\documentclass{aa}

\usepackage{multirow}
\usepackage{graphicx}
\usepackage{tabularx}
\usepackage{txfonts}
\usepackage{hyperref}
\usepackage{booktabs}
\usepackage{verbatim} 
 
\hypersetup{
    colorlinks=true,
    linkcolor=blue,
    filecolor=magenta,      
    urlcolor=blue,
    pdftitle={Overleaf Example},
    pdfpagemode=FullScreen,
    citecolor=blue
    }
\newcommand\code[1]{\textsc{\MakeLowercase{#1}}}
\newcommand{\ra}[1]{\renewcommand{\arraystretch}{#1}}
\newcommand{\bba}{$^{\scriptstyle 3\mathrm{D}}$B{\sc arolo}}

\def\CII{\hbox{[C~$\scriptstyle\rm II $]}}
\def\msolar{{\rm M}_{\odot}}

\newcommand{\quotes}[1]{``#1''}

\begin{document}

   \title{Dynamical characterization of galaxies up to $z \sim 7$}

   \subtitle{} 

   \author{F. Rizzo
        \inst{1, 2},
        M. Kohandel\inst{3}
         \and
        A. Pallottini\inst{3}
        \and
        A. Zanella\inst{4}
        \and
        A. Ferrara\inst{3}
        \and
        L. Vallini\inst{3}
        \and
        S. Toft\inst{1, 2}
          }

   \institute{Cosmic Dawn Center (DAWN)
   \and Niels Bohr Institute, University of Copenhagen, Jagtvej 128, 2200 Copenhagen N, Denmark\\
              \email{francesca.rizzo@nbi.ku.dk}
              \and
             Scuola Normale Superiore, Piazza dei Cavalieri 7, I-56126 Pisa, Italy
             \and 
             Istituto Nazionale di Astrofisica, Vicolo dell'Osservatorio 5, I-35122 Padova, Italy
             }

   \date{}

 
  \abstract
   {The characterization of the dynamical state of galaxies up to $z \sim 7$ is crucial for constraining the mechanisms that drive the mass assembly in the early Universe. However, it is unclear whether the data quality of typical observations obtained with current and future facilities is sufficient to perform a solid dynamical analysis at these redshifts.}
   {This paper defines the angular resolution and signal-to-noise ratio (S/N) required for a robust characterization of the dynamical state of galaxies up to the Epoch of Reionization. The final aim is to help design future spatially-resolved surveys targeting emission lines of primeval galaxies.}
   {We investigate the [CII]-158$\mu$m emission from $z \sim 6-7$ Lyman Break Galaxies from the \code{SERRA} zoom-in cosmological simulation suite. The \code{SERRA} galaxies cover a range of dynamical states: from isolated disks to major mergers. We create mock observations with various data quality and apply the kinematic classification methods commonly used in the literature. These tests allow us to quantify the performances of the classification methods as a function of angular resolution and S/N.}
   {We find that barely-resolved observations, typical of line detection surveys, do not allow the correct characterization of the dynamical stage of a galaxy, resulting in the misclassification of all disks in our sample. However, even when using spatially-resolved observations with data quality typical of high-$z$ galaxies (S/N $\sim 10$, and $\sim 3$ independent resolution elements along the major axis), the standard kinematic classification methods, based on the analysis of the moment maps, fail to distinguish a merger from a disk. The high angular resolution and S/N needed to correctly classify disks with these standard methods can only be achieved with current instrumentation for a handful of bright galaxies.
   We propose a new classification method, called PVsplit, that quantifies the asymmetries and morphological features in position-velocity diagrams using three empirical parameters. We test PVsplit on mock data created from \code{SERRA} galaxies, and show that PVsplit can predict whether a galaxy is a disk or a merger provided that S/N $\gtrsim 10$, and the major axis is covered by $\gtrsim 3$ independent resolution elements.
   }
   {}


   \maketitle
%
\section{Introduction}
In the current paradigm of cosmological structure formation, the evolution of galaxies results from the interplay between different processes: accretion of cold gas, minor and major mergers, stellar and active-galactic-nuclei feedback \citep[e.g.,][]{Somerville_2015, Cimatti_2019}. Some of these processes (e.g., gas accretion) lead to the formation of a disk structure \citep[e.g.,][]{Dekel_2009, Pallottini_2017dahlia, Kohandel_2019, Kretschmer_2021, Tamfal_2021} or the increase of the gas turbulence within early galaxies \citep[e.g., stellar feedback, minor mergers, ][]{Ceverino_2015, Danovich_2015, Pillepich_2019, Kohandel_2020, Kretschmer_2021}. On the contrary, frequent major merger events or counter-rotating gas streams can be potentially disruptive, destroying and preventing the rebuilding of the disks on timescales comparable or longer than the dynamical times of the galaxy \citep[e.g.,][]{Bournaud_2007, Dubois_2012, Zolotov_2015, Dekel_2020}. Studying the dynamics of star-forming galaxies across cosmic time is crucial to constrain the relative importance of processes mostly contributing to the growth of galaxies and the evolution of their morphology \citep[e.g.,][]{Wisnioski_2015, Turner_2017, Johnson_2018, Krumholz_2018, Wisnioski_2018, Kohandel_2020, Ejdetjarn_2021, Romano_2021}. 

The prevalence of rotating disks among star-forming galaxies at $z \sim$ 1 - 3 has supported the idea that, during this epoch, galaxies mainly grow in mass due to smooth accretion of cold gas inflowing through the cosmic web (i.e., secular growth), while mergers seem to play only a minor role \citep{Wisnioski_2015}. However, the fraction of rotating disks at these intermediate redshifts has been debated in the last years. At $z \sim 1$, the fraction of disks in the star-forming galaxy population goes from 80$\%$ \citep{Wisnioski_2015, Wisnioski_2018} to 42$\%$ \citep{Rodrigues_2017} depending on the criteria employed for the classification. \citet{Simons_2019} showed that the angular resolutions of current near-infrared Integral-Field-Unit (IFU) observations hampered the possibility to robustly distinguish the regular rotation of a disk from the orbital motions of interacting galaxies. By analyzing synthetic IFU observations of simulated galaxies at $z \sim $2, \citet{Simons_2019} concluded that the probability that a merger is classified as a disk could be as high as 100$\%$ for close-pair merging galaxies. 

The characterization of the dynamics of galaxies just after (4 $\lesssim z \lesssim$ 6) and within the Epoch of Reionization ($z \gtrsim 6$) is still in its infancy as spatially-resolved observations targeting emission lines are available only for about ten of sub-mm sources \citep[e.g.,]{Rizzo_2020, Lelli_2021, Rizzo_2021} and a handful of Lyman-Break galaxies \citep[e.g.,][]{Jones_2017, Fujimoto_2021}. Instead, the number of marginally-resolved observations aiming at inferring the integrated properties of early galaxies has increased in the last five years \citep{Smit_2018, LeFevre_2020, Bouwens_2021}. In their pioneering work, \citet{Smit_2018} used marginally-resolved Atacama Large Millimeter/Submillimetre Array \citep[ALMA,][]{Wootten_2009} observations of the [CII]-158$\mu$m emission line to show that two star-forming galaxies at $z \sim 7$ have smooth velocity gradients and interpreted them as being rotationally-supported disks. Similar results have been obtained in studies of individual galaxies at $z \sim$ 6 - 8 \citep{Bakx_2020, Harikane_2020}. However, given the low angular resolution of these data, they could not rule out the possibility that one or more merging [CII]-bright satellites were mimicking the smooth gradient observed in the velocity maps \citep[see discussion in][]{Smit_2018, Simons_2019}. Instead, a gradient in the velocity fields of two Lyman-Break galaxies at $z = 6.1$ and 7.1, combined with the identification of two compact components, has been interpreted as evidence of mergers \citep{Jones_2017, Hashimoto_2019}. Recently, \citet{LeFevre_2020} and \citet{ Romano_2021} performed the first systematic morpho-kinematic analysis of a statistically significant sample of $z \sim$ 4 - 6 main-sequence galaxies from the ALMA Large Program to INvestigate [CII] at Early times (ALPINE) survey. After combining the morphological and kinematic analysis of the [CII] observations with the rest-frame UV and optical data, \citet{Romano_2021} concluded that 23 out of the 75 ALPINE galaxies are merging systems. This large fraction of mergers in the ALPINE sample could imply a significant contribution of major mergers to the mass assembly in the early Universe \citep{LeFevre_2020, Romano_2021}, confirming previous results based on the count of close-pair galaxies in photometric surveys \citep{Mantha_2018, Duncan_2019}. However, due to the sensitivity and angular resolutions of the ALPINE observations, the kinematic characterization and, consequently, the merger fraction may be uncertain.

To summarize, both at intermediate and high-$z$, characterizing the kinematic state of galaxies has been challenging \citep[e.g.,][]{Goncalves_2010, Smit_2018, Simons_2019} since the classification techniques used in the literature \citep[e.g.,][]{Shapiro_2008, Bellocchi_2012, Wisnioski_2015, Wisnioski_2018} have focused on the analysis of the kinematic maps (velocity and velocity-dispersion fields). At the typical angular resolution of current observations, any irregularities and disturbances in the kinematic maps of distant galaxies are usually smoothed out so that galaxies appear more regular than they are \citep[e.g.,][]{Bellocchi_2012, Simons_2019}.
Further, kinematic maps cannot be considered a thorough representation of the data in their native three-dimensional space (RA, Dec, Frequency) since they are obtained after masking the low signal-to-noise (S/N) regions and integrating them along the spectral axis. In addition, most of the kinematic classification techniques have been tested and calibrated on IFU observations at intermediate-$z$ \citep[e.g.,][]{Shapiro_2008, Bellocchi_2012, Wisnioski_2015, Simons_2019}. A quantitative investigation of their range of applicability to interferometric observations with a wide range of S/N ratios and angular resolutions is still missing  \citep[e.g., see also discussion in][]{Glazebrook_2013}. 

With this work, we aim at filling this gap by quantifying how the data quality affects the correct characterization of the dynamical stage of high-$z$ galaxies. The results of this paper will help to design future surveys aimed at constraining the dynamical properties of galaxies up to the Epoch of Reionization (EoR). To this end, we generate mock ALMA observations \citep{Kohandel_2020} by using simulated galaxies from the \code{SERRA} suite at $z \sim$ 6 - 7 \citep{Pallottini_2022}.
After analyzing the mock data sets as we would analyze real data, we discuss the potential biases of the typical kinematic classification methods as a function of the angular resolutions and the S/N ratios of the observations. In this first paper, we also present and discuss a new technique aiming at distinguishing different dynamical states of galaxies even at low-angular resolutions. Detailed analysis of a representative sample of galaxies will be presented in the second paper of this series.

This paper is organized as follows: we summarize the main characteristics of the zoom-in cosmological simulations adopted in this work and the galaxies selected for our analysis in Sects. \ref{sec:simulation_serra} and \ref{sec:simulated_data}. In Sects. \ref{sec:mock_data} and \ref{sec:barolo}, we describe how we created the mock data and extracted the kinematic properties. The application of the standard kinematic classification techniques is presented and discussed in Sect. \ref{sec:classification}. In Sect. \ref{sec:newclas}, we introduce a new proxy for characterizing different dynamical states of galaxies using medium and low-angular resolution observations. Finally, we summarize the results in Sect. \ref{sec:conclusions}.

\section{Numerical simulations}\label{sec:simulation_serra}

\code{SERRA} is a suite of simulations that focuses on zooming in on the formation and evolution of galaxies in the EoR \citep[see][for details]{Pallottini_2022}. Gas and dark matter are evolved with a customized version of adaptive mesh refinement code \code{RAMSES} \citep{Teyssier_2002}.
Adopting \code{KROME} \citep{Grassi_2014}, \code{SERRA} includes non-equilibrium chemical (and thermal) evolution up to the formation of H$_2$ \citep{Pallottini_2017}. The chemical network used in \code{SERRA} includes $\rm{H}$, $\rm{He}$, $\rm{H}^+$, $\rm{H}^-$, $\rm{He}$, $\rm{He}^+$, $\rm{He}^{++}$, $\rm{H}_2$, $\rm{H}_2^{+}$ and electrons \citep{bovino:2016}. Metals and dust contribute to the cooling of the gas and H$_2$ formation.
Metallicity is tracked as the sum of heavy elements, assuming solar abundance ratios of different metal species \citep{Asplund_2009}. Dust formation and evolution are not tracked explicitly: it is assumed that the dust-to-gas mass ratio scales with metallicity, and an MW-like grain size distribution is adopted.

Molecular hydrogen is converted into stars following a Kennicutt-Schmidt-like relation \citep{Schmidt_1959, Kennicutt_1998}.
Stars in \code{SERRA} affect the chemical and energy budget of the interstellar medium (ISM) via feedback processes that include Type II and Ia Supernov\ae, winds from OB and AGB stars, and radiation pressure, which give both a thermal and turbulent energy contribution to the ISM \citep{Agertz_2013,Pallottini_2017dahlia}.
Further, stars contribute to the interstellar radiation field (ISRF), which is tracked on the fly using the radiative transfer module \code{RAMSES-RT} \citep{Rosdahl_2013}, which is consistently coupled to the chemical evolution of the gas \citep{Pallottini_2019, Decataldo_2019}.

In the \code{SERRA} suite, simulations evolve from cosmological initial conditions, computed with \code{MUSIC} \citep{Hahn_2011} at $z=100$. In each simulation the comoving volume is $(20\,{\rm Mpc}/{\rm h})^{3}$ and its evolution is followed down to $z=6$. For each independent volume, the simulation zoom-in on the formation and evolution of $\sim 10-20$ galaxies, whose ISM is resolved on scales of $\simeq 1.2\times 10^4 \msolar$, $\simeq 30\, \rm pc$, i.e. masses and sizes typical of Molecular Clouds \citep[e.g.][]{Federrath_2013}. Currently, at $z\simeq 6-7$ the \code{serra} suite has a total of $\simeq 200$ galaxies with mass range $\sim 10^{8}-10^{10} \msolar$ \citep{Pallottini_2022}.

Individual ion abundances (such as $\rm{C}^+$) and line emissions (such as \CII) are recomputed in post-processing by using the spectral synthesis code \code{CLOUDY} \citep{Ferland_2017}. The [CII] emission line is known to originate from all the gas phases of the ISM, albeit predominantly from cold neutral/molecular gas in the so-called photodissociation regions \citep[see][for a recent review]{Wolfire_2022}, thus is a unique tracer of the total gas mass in galaxies.
In \code{SERRA} the internal structure of molecular clouds is not resolved; thus, an additional model is adopted to account for the turbulent and clumpy structure of the ISM \citep[see][for details]{Vallini_2018,Pallottini_2019}.
Further modeling is performed to bridge the outputs of zoom-in simulations and IFU observations (such as ALMA) by generating Hyper-spectral Data Cubes (HDC) for each given emission line \citep[see][for the details]{Kohandel_2020}.
To summarize, we compute \CII~luminosity on a cell by cell basis depending on gas density, metallicity, thermal+turbulent broadening, column density, and ISRF intensity; then, given the gas position, line of sight velocity, and thermal+turbulent broadening, we construct HDCs with two spatial and one spectral dimension.
Such data products are especially handy for galaxy kinematic studies and can be used for direct comparison of the simulations and spatially-resolved galaxy observations.

In this work, we select cubic regions centered on chosen galaxies with a fixed inclination adopting a side-length of $8\,\rm{kpc}$ and mapping the volume to \CII~emission HDCs with $256^3$ voxels, i.e. a setup similar to \citet{Kohandel_2020}.
Thus, our HDCs have a spectral resolution of $ \simeq 3.1\,\rm{km\,s}^{-1}$ and a spatial resolution of $ \simeq 31.2 \, \rm{pc}$, corresponding to an angular resolution of $0.005^{\prime\prime}$ at $z=6$.

\section{Sample selection}\label{sec:simulated_data}

\begin{figure*}
        \centering
       \includegraphics[width=0.99\textwidth]{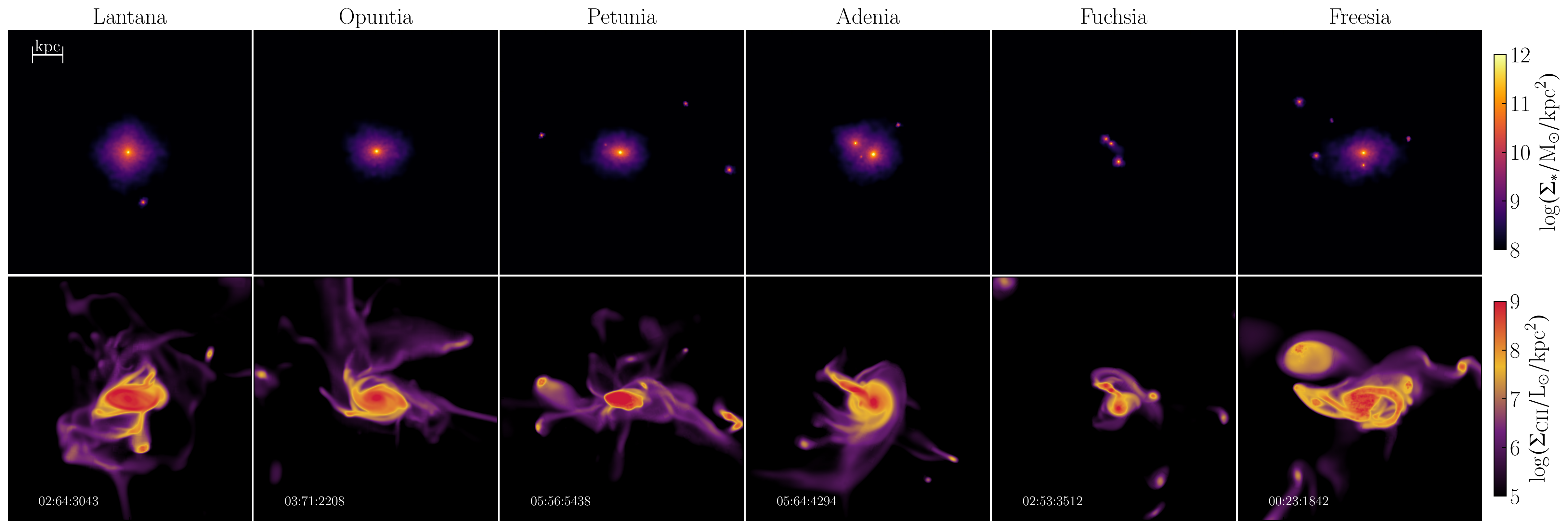}
        \caption{Sample of \code{SERRA} galaxies adopted in this work. For each galaxy we plot the the stellar surface density ($\Sigma_\star$, \textit{upper panels}) and the \CII~line surface brightness ($\Sigma_{[\rm{CII}]}$, \textit{lower panels}) for a viewing angle of 60 degrees. The field of view has a size of 8 kpc. See Tab. \ref{tab:galaxy_overview} for the main properties of the galaxies.
        \label{fig:visualize_sample}
        }
\end{figure*}
\begin{figure}
        \centering
        \includegraphics[width=0.49\textwidth]{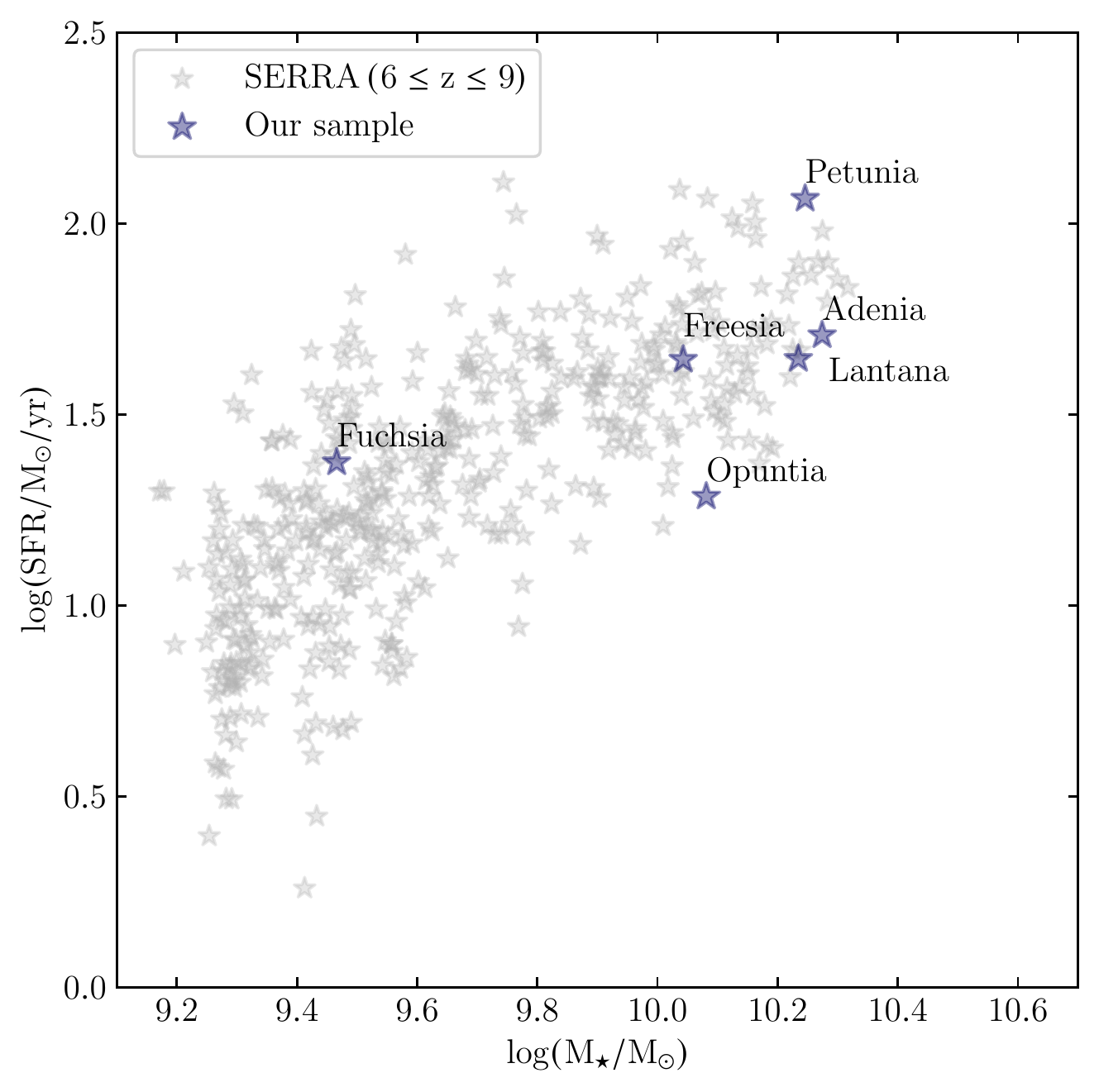}
        \caption{The position of galaxies in our sample on SFR-$M_\star$ plane. The gray stars are a sample of EoR galaxies from the \code{serra} suite \citep[see][]{Pallottini_2022,kohandel:2022}.\label{fig:sfr-mstar}}%
\end{figure}

\begin{table}
\caption{Properties (stellar mass $M_\star$, gas mass $M_g$, star formation rate SFR and \CII~line luminosity $L_{\rm{[CII]}}$) of \code{SERRA} galaxies analyzed in this work \citep[see][]{Pallottini_2022,kohandel:2022}. Opuntia, Lantana and Petunia are the three disk galaxies, Freesia is the disturbed disk, and Adenia and Fuchsia are the two mergers.  
\label{tab:galaxy_overview}
}
\begin{center}
\begin{tabular}{ccccccc}
\hline
Name & $z$ & $\rm{M}_{\star}$& $\rm{M}_{\rm{g}}$ & $\rm{SFR}$ & $\rm{L}_{\rm{[CII]}}$ \\ 
 ~ & ~ & $[10^{10}\msolar]$ & $[10^{9}\msolar]$ & $[\msolar \rm{yr}^{-1}]$ & $[10^{8}\rm{L}_{\odot}]$ \\
\hline
Opuntia & $6.1$ & $1.2$ & $4.5$ & $19$  & $3.1$ \\  
Lantana & $6.4$ & $1.7$ & $7$   & $44$  & $6.6$ \\  
Petunia & $7.0$ & $1.8$ & $2.8$ & $116$ & $5.1$ \\  
Freesia & $6.5$ & $1.1$ & $3.8$ & $44$  & $5.0$ \\  
Adenia  & $6.5$ & $1.9$ & $5.4$ & $51$  & $3.4$ \\  
Fuchsia & $7.2$ & $0.3$ & $1.7$ & $24$  & $1.0$ \\  
\hline
\end{tabular}
\end{center}
\end{table}

Our sample contains 6 galaxies in the redshift range $6 \lesssim z \lesssim 7$ from the \code{SERRA} suite; we select 3 rotating disks (Lantana, Opuntia and Petunia), 1 disturbed disk (Freesia) and 2 major mergers\footnote{The mass ratio between the merging companions is $>0.4$.} (Adenia and Fuchsia).
Similarly to \citet{Kohandel_2019}, the dynamical stages of these galaxies are identified based on the morphology of the \CII~line surface brightness maps and the corresponding spectra extracted for their face-on \footnote{We call a galaxy face-on when we orientate the l.o.s. parallel to the eigenvector of the inertia tensor of the gas mass distribution with the largest eigenvalue.} and edge-on views.
Our sample spans a stellar mass range $\rm{M}_{\star} \sim (0.3-1.7)\times 10^{10}\,M_\odot$, a star-formation rate SFR $\sim 20-100\,M_\odot \rm{yr}^{-1}$ and \CII~luminosity range $\rm{L}_{\rm{[CII]}} \sim (0.6-6.3)\times 10^{8}\,L_\odot$. In Tab. \ref{tab:galaxy_overview}, general properties of these galaxies are tabulated. In Fig. \ref{fig:sfr-mstar}, the position of galaxies in our sample is shown on the SFR-$M_\star$ plane with respect to a large sample of EoR galaxies from \code{serra} presented in Kohandel et al. in prep. Galaxies in our sample have a specific star formation rate $1.6\,\rm{Gyr}^{-1}\le \rm{sSFR}\equiv \rm{SFR}/M_\star\le 8 \,\rm{Gyr}^{-1}$ with Opuntia having the lowest value and Fuchsia having the highest value. 

For each galaxy, HDCs are generated using three viewing angles (30, 60, and 80 $\deg$).
To have a visual overview, in Fig. \ref{fig:visualize_sample}, the stellar surface density ($\Sigma_\star$) and \CII~line surface brightness ($\Sigma_{[\rm{CII}]}$) maps for the viewing angle of 60$\deg$ of our sample are plotted.
     
\section{Mock observations}\label{sec:mock_data}

To study how the data quality (angular resolution and S/N) impacts the classification of disks and mergers, we generated 102 mock observations from the simulated \CII~emission HDCs of galaxies described in Sect. \ref{sec:simulated_data}.
We rebin the HDC to have data cubes with spectral channels of width 30 km\,s$^{-1}$, as typically done for observations of galaxies at $z \gtrsim 4$ \citep[e.g.,][]{Lelli_2021, Rizzo_2021}. This spectral resolution allows us to recover the velocity dispersion of \code{SERRA} galaxies with typical values $\gtrsim 30$ km\,s$^{-1}$ \citep{Kohandel_2020}. Based on the data quality, we divided the synthetic observations into the following five categories (see Table \ref{table:mock}):
\begin{itemize}
    \item[a)] \textbf{Ideal data:} ideal observations at high angular resolution and very high S/N. These mock data are obtained after convolving the cubes with a Gaussian point-spread-function with a full width at half maximum of 0.02" ($\simeq 120\, \rm pc$). Then, we added Gaussian noise to each pixel to reach an S/N ratio of $\sim$ 40 \footnote{The S/N is computed as the median of the maximum S/N in each spectral channel.}.\label{list:mock0}
    \item[b)] \textbf{ALMA high-angular resolution, high S/N data:} We used the tasks \code{SIMOBSERVE} in the \code{CASA} package \citep{McMullin_2007} to create ALMA interferometric observations at the nominal angular resolution of 0.02" ($\simeq 120\, \rm pc$) and with a S/N ratio of 10. The galaxies are then imaged with a natural weighting of the visibilities using the task \code{SIMANALYZE}. On average, the area of the galaxies with [CII] surface brightness larger than 3 times the r.m.s noise (3\,rms) is covered by $\approx 10$ resolution elements in each spectral channel. \label{list:resolvedalma}
    \item[c)] \textbf{ALMA medium- and low-angular resolution, high S/N data:} Similar to point b), but with nominal angular resolutions of 0.05" ($\simeq 0.3\, \rm kpc$) and 0.1" ($\simeq 0.6\, \rm kpc$). The area of the galaxies at $\gtrsim$ 3\,rms is covered on average by 3 and 2 resolution elements per spectral channel for the medium and low-resolution data, respectively.
    \item[d)] \textbf{ALMA high- and medium-angular resolution, low S/N data:} For the galaxies at $i \sim 60 \deg$, we created ALMA mock data using the same methodology described at point b) but with an S/N of $\sim 5$. In this case, the area of the galaxies with [CII] surface brightness $\gtrsim$ 3\,rms is covered by $\approx 4$  and 2 resolution elements in each spectral channel for the high and medium-resolution data, respectively.\label{list:mediumres}
    \item[e)] \textbf{ALMA barely-resolved, low S/N data:} We used \code{SIMOBSERVE} and \code{SIMANALYZE} to create mock data matching the data quality of the ALPINE sample as well as most observations of $z \gtrsim 4$ galaxies \citep[e.g.,][]{LeFevre_2020, Smit_2018, Bakx_2020}. On average, the ratios between the deconvolved [CII] effective radius and the beam size range between 0.3 \citep[e.g., ALPINE sample][]{LeFevre_2020, Fujimoto_2020} and 0.5 \citep[e.g.,][]{Smit_2018, Bakx_2020}. The CII~emitting gas in \code{SERRA} galaxies described in Sect. \ref{sec:simulated_data} has an average effective radius of $\sim 0.07"$. For this dataset a resolution of 0.15" ($\simeq 0.9\, \rm kpc$) allowed us to reproduce the typical angular resolutions of current $z \gtrsim 4$ observations. The barely-resolved ALMA data have an S/N ratio of $\sim$ 4 that corresponds to S/N ratios in the range 6 to 10 in the moment-0 maps, comparable with the values reported in \citet{Bethermin_2020} and \citet{Fujimoto_2020} for the ALPINE galaxies. On average, the area of the galaxies with [CII] surface brightness $\gtrsim$ 3\,rms is covered by $\approx 0.3$ resolution elements in each spectral channels.\label{list:alpine}
\end{itemize}
    \begin{table*}\centering
    \caption{Summary of the mock observations created for our sample of 6 galaxies (see Sect. \ref{sec:mock_data} for further details).}
    \ra{1.3}
    \begin{tabular}{@{}cccccccccc@{}}\toprule
    \multicolumn{1}{c}{} & \phantom{abc} & \multicolumn{1}{c}{Ideal Mock} &
    \phantom{abc} & \multicolumn{6}{c}{ALMA Mocks} \\ \cmidrule{3-3} \cmidrule{5-10}
        && High S/N  &&  High S/N &  \multicolumn{2}{c}{High S/N}  & \multicolumn{2}{c}{Low S/N } & Low S/N\\ 
        && High res &&  High res &  \multicolumn{2}{c}{Medium \& low res}  & \multicolumn{2}{c}{High \& medium res} & Barely resolved\\ \midrule\midrule
    S/N        && 40    && 10    & 10    & 10   & 5     & 5     & 4    \\\noalign{\smallskip}
    Resolution && 0.02" && 0.02" & 0.05" & 0.1" & 0.02" & 0.05" & 0.15"\\\noalign{\smallskip}
    \bottomrule
    \end{tabular}
    \label{table:mock}
    \end{table*}

\begin{figure*}
        \centering
        \includegraphics[width=0.99\textwidth]{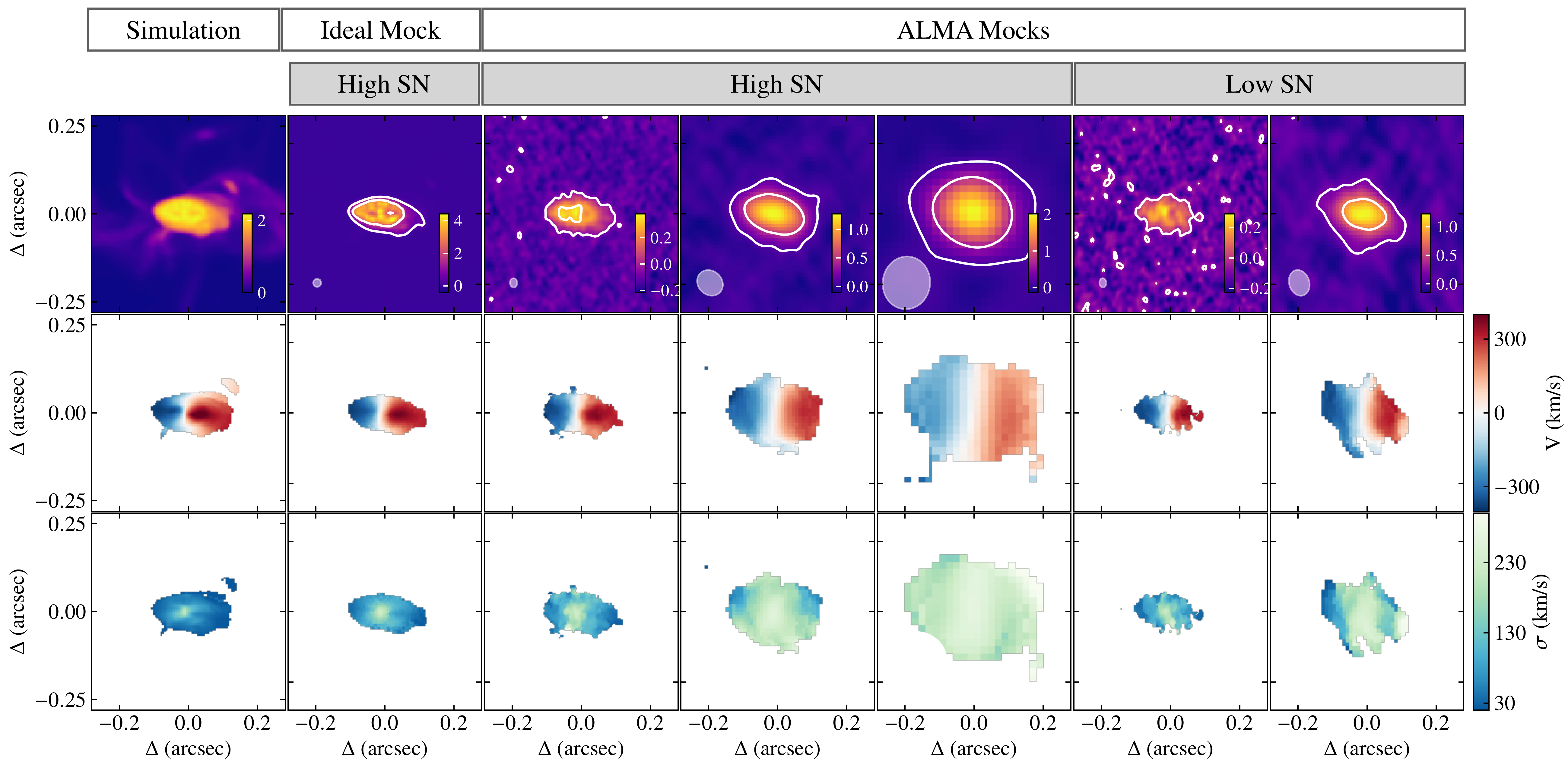}
        \caption{The three rows show the moment-0, 1, and 2 for the disk Petunia at 60$\deg$ obtained from the simulated cube (column 1) and the mock cubes described at points a) - d) in Sect. \ref{sec:mock_data} (columns 2 to 7) at different angular resolutions (the beam is shown in the bottom left corner as the gray ellipse). In the mock moment-0 maps, the white contours are at 2 and 10 times the r.m.s. noise and the color bars are in units of 10 mJy/pixel km s$^{-1}$ in the first column and 100 mJy/beam km s$^{-1}$ for the mock observations.
        \label{fig:5438_moments}%
        }
\end{figure*}

\begin{figure*}
        \centering
        \includegraphics[width=0.99\textwidth]{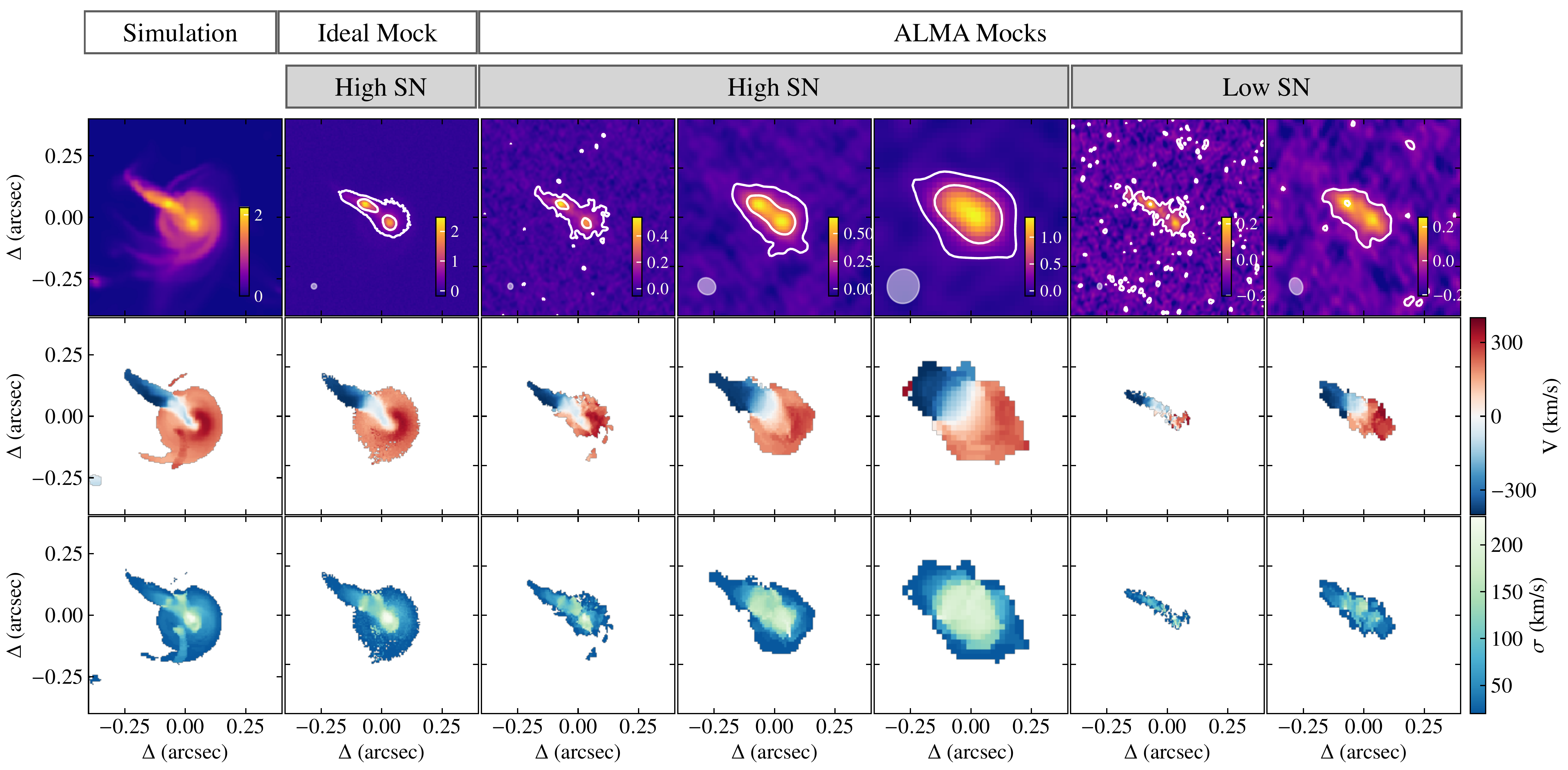}
        \caption{Same as in Fig. \ref{fig:5438_moments} but for the merging system Adenia at 60$\deg$.
        \label{fig:4294_moments}%
        }
\end{figure*}

Figs. \ref{fig:5438_moments} and \ref{fig:4294_moments} show the moment-0 ([CII] integrated across the spectral axis), moment-1 (flux-weighted velocity) and moment-2 (flux-weighted velocity dispersion) maps obtained from the mock data a) - d) for two representative galaxies, the disk Petunia and the merger Adenia at $i = 60 \deg$.

We note that currently only a handful of observations of lensed and non-lensed dusty star-forming galaxies at $z \gtrsim 4$ have data quality comparable to that described at points b)-d) \citep[e.g.,][]{Neeleman_2020, Rizzo_2020, Fraternali_2021, Lelli_2021, Rizzo_2021}.
Most observations of main-sequence galaxies at $z \gtrsim 4$ \citep[e.g.,][]{LeFevre_2020, Smit_2018, Bakx_2020} have been designed, instead, for line detection, and therefore their data quality is comparable to the 18 mock data described at point e). A test for this low-resolution and low-S/N data is discussed in Sect. \ref{sec:alpine}. However, for the rest of the paper, we adopt the 84 synthetic observations described at points a)-d), if not otherwise stated, as in the near future, medium- and low-angular resolution, high S/N ALMA, and JWST observations will be available for tens $z \gtrsim 4$ galaxies.

\section{Kinematic modeling}\label{sec:barolo}

We assumed that the \CII~emitting gas moves in circular orbits to extract the kinematic properties of galaxies in our sample. Deviations from pure circular orbits include radial and vertical motions due to outflows and inflows or asymmetry of the galactic gravitational potential driven by mergers, bars, or spiral arms \citep[e.g.][]{Fraternali_2001, Almeida_2021, DiTeodoro_2021, Rizzo_2021, Yttergren_2021}. %
Such deviations from circular motions are evident in the merging galaxies of our sample, especially at high- and medium-angular resolutions observations (see Fig. \ref{fig:4294_moments} as a representative example).  However, when considering low-resolution observations, these merging systems look similar to rotating disks in the moment-1 maps. Therefore, we decide to fit all the galaxies in our sample assuming a simple rotating disk \footnote{Note that for the mergers, the resulting kinematic parameters (e.g., rotation velocity and velocity dispersions) do not have a physical meaning and thus can not be used to infer physical parameters of the galaxies (e.g., virial mass, dark matter halo mass)}.

Kinematic measurements of high-$z$ galaxies are very challenging, even in the case of disk galaxies. Complications arise due to the effect of beam smearing, which artificially decreases the rotation velocity and increases the velocity dispersion measurements \citep{Bosma_1978, Swaters_1999, DiTeodoro_2015, Kohandel_2020}.
To recover the intrinsic values of rotation velocity and velocity dispersion, one should adequately account for the beam-smearing effect using either a-posteriori or forward modeling techniques.
In the first case, the velocity dispersions and rotation velocities are derived from the 2D maps (moment-1 and 2) and then corrected a-posteriori using analytical functions that take into account the velocity gradients \citep[e.g.][]{Swinbank_2012, Stott_2016, Federrath_2017} or the sizes of the galaxies and the beam \citep[e.g.][]{Burkert_2016, Johnson_2018}. In the second case, the data are fitted directly in their native three-dimensional (3D) space using a 3D model convolved with the beam of the observations \citep[e.g.,][]{DiTeodoro_2015, Bouche_2015}.

In this paper, we use the forward-modeling technique through the code \bba\ \citep{DiTeodoro_2015} to characterize the kinematic properties of our mock galaxies. \bba\ creates 3D realizations of a tilted-ring model \citep{Rogstad_1974}: we assume the galaxy to be a disk divided into a series of concentric circular rings, each with its kinematic (i.e., systemic velocity $V_{\mathrm{sys}}$, rotation velocity $V_{\mathrm{rot}}$ and velocity dispersion $\sigma$) and geometric properties (i.e., center, inclination angle $i$ and position angle).

Under the assumption that the galaxy is a thin disk with negligible non-circular motions, the line-of-sight velocity $V_{\mathrm{los}}$ at a radius $R$ is given by 
\begin{equation}
V_{\mathrm{los}} = V_{\mathrm{sys}}+V_{\mathrm{rot}}(R)\cos\phi \sin i,
\end{equation}
where $\phi$ is the azimuthal angle in the disk plane. After producing the model disk, \bba\ convolves it with the beam of the observations, and then it calculates and minimizes the residuals between the data and the model. For disk galaxies, this methodology allows for a robust recovery of the rotation velocity and velocity dispersion profiles since it largely mitigates the effects of beam smearing \citep{Bosma_1978, Swaters_1999, DiTeodoro_2015}.
Further details about the assumptions made to fit our mock data are described and discussed in Appendix \ref{appendix:kinematic_assumptions}.

For the rest of the paper, the kinematic properties recovered for our sample are shown as a function of the number of independent resolution elements $N_{\mathrm{IRE}}$. The latter is calculated as  $R_{\mathrm{max}}/\sqrt{B_{\mathrm{maj}} B_{\mathrm{min}}}$, where $R_{\mathrm{max}}$ is the external radius of the last ring of the kinematic model and $B_{\mathrm{maj}}$ and $B_{\mathrm{min}}$ are the sizes of the beam along the major and minor axis, respectively. The values of $N_{\mathrm{IRE}}$ for our mock data are listed in Table \ref{table:nire}.
We note that the kinematic fitting with \bba\ is not feasible for the 18 barely-resolved mock data sets since the number of pixels at 2\,rms is not sufficient to define the ring regions and constrain the model.

\subsection{Kinematic outputs}\label{sec:kinematic_outputs}

In addition to the best-fit values for the geometrical and kinematic parameters at each radius $R$, the outputs from \bba\ are the following:
\begin{enumerate}
    \item model cubes convolved with the same beam of the data.
    \item moment-0, 1, and 2 maps of the data and model.
    \item position-velocity (PV) diagrams\footnote{The PV diagrams are slices extracted from data cubes along a specific spatial direction (e.g., the kinematic major axis).} extracted along the major and minor-axis from both the data and the model cubes. 
\end{enumerate}
For a rotating disk, the major-axis PV diagram has a characteristic s-shape, and the minor-axis PV diagram is symmetric with respect to the axes defining the systemic velocity and the center. The presence of non-circular motions (due to gas inflowing or outflowing) can be identified by asymmetries in the minor-axis PV diagrams or any emission in the so-called forbidden regions (i.e., forbidden for rotation), that is, the quadrants not occupied by the s-shape \citep[e.g., see discussion in][]{Fraternali_2002, DiTeodoro_2021}.

In the following, we describe the comparison between the data and the kinematic models for the disks and mergers of our sample.

\begin{figure*}
        \centering
        \includegraphics[width=0.99\textwidth]{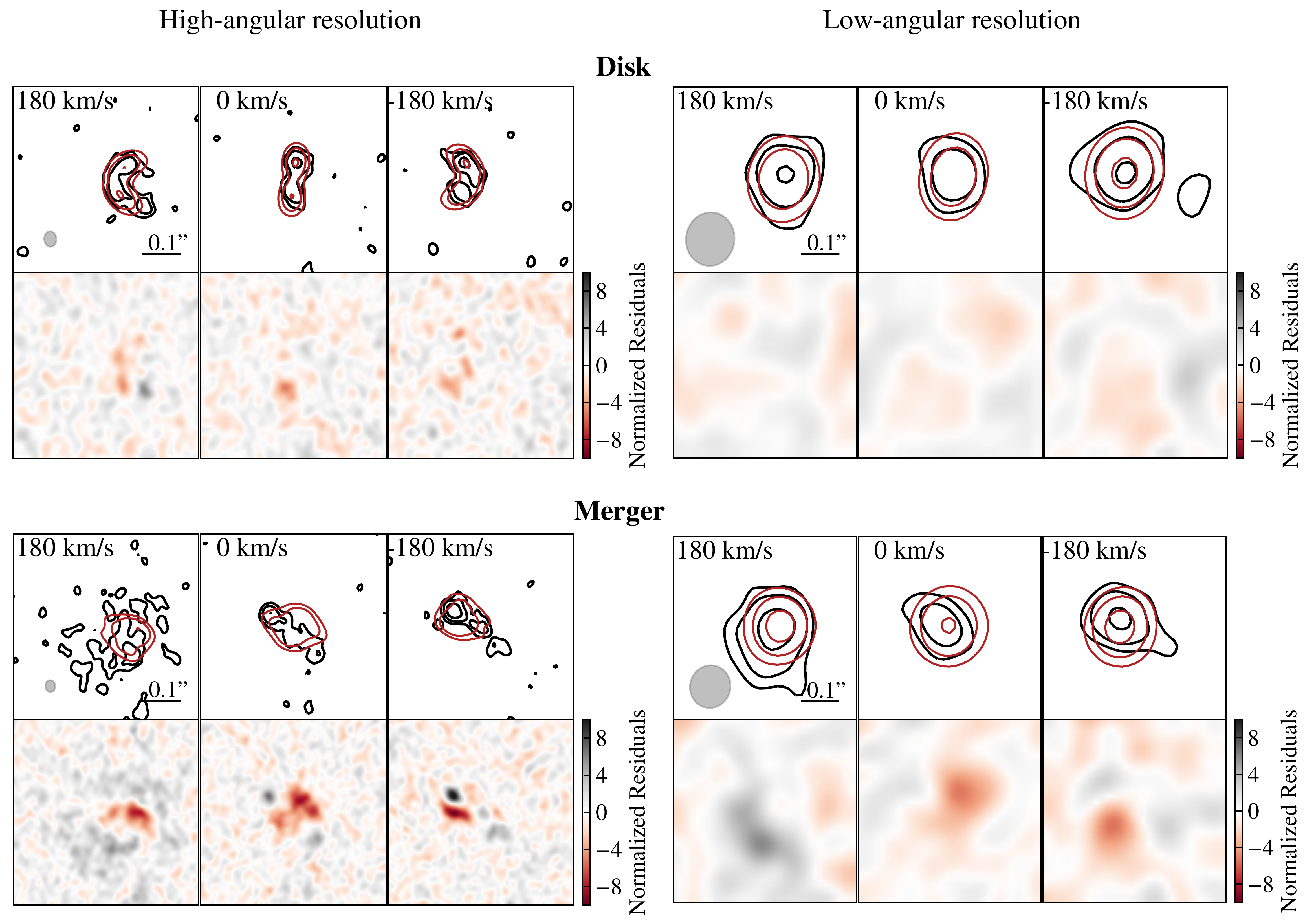}
        \caption{Three representative spectral channels at velocities 180, 0, -180 km s$^{-1}$, for the disk Petunia (upper panels) and the merger Adenia (bottom panels) at 60 $\deg$. The left and right subfigures show the high S/N ALMA mock data at high and low-angular resolution. For each subfigure: the upper row shows the data (black contours) and the \bba\ model (red contours). The contour levels are at 2.5, 5, 10 times the r.m.s. noise per channel. The gray ellipses in the first panels denote the synthetized beam of the observations. The bottom row shows the residuals normalized to the r.m.s. noise.   }
        \label{fig:5438_ch}%
\end{figure*}

\begin{figure*}
        \centering
        \includegraphics[width=0.99\textwidth]{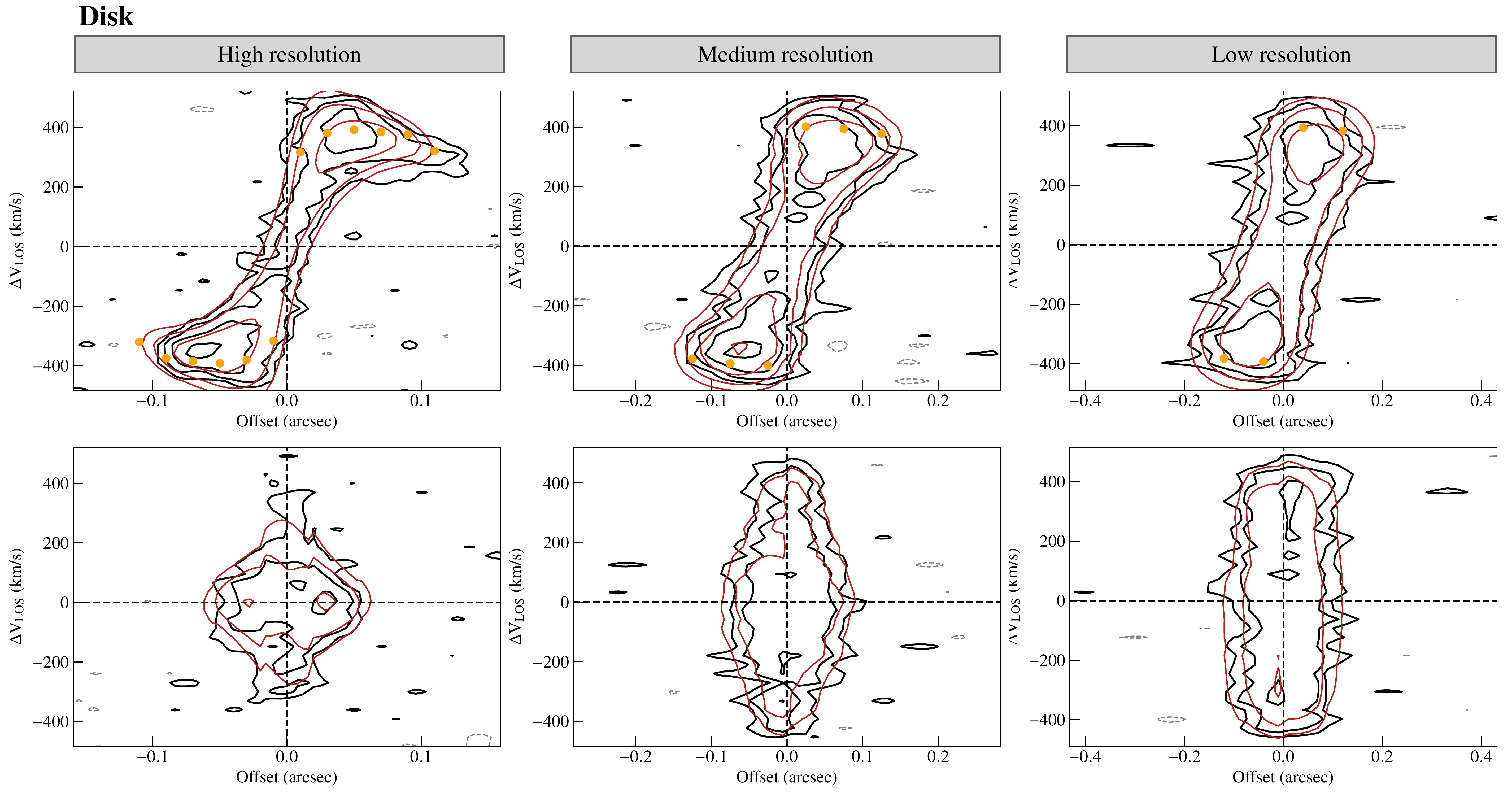}
        \caption{Position-velocity diagrams along the kinematic major (upper panels) and minor axis (bottom panels) for the disk Petunia at 60 $\deg$. The horizontal axis shows the offset from the galaxy center, and the vertical axis represents the line-of-sight velocity centred at the systemic velocity of the galaxy.
        From left to right, the high S/N ALMA mock data at high-, medium- and low-angular resolutions are shown. The black and red contours show the data and the \bba\ model, respectively. The level contours are at -2.5, 2.5, 5 and 10 times the r.m.s. noise. The orange circles show the best-fit rotation velocities (not corrected for inclination) derived using \bba\ .}
        \label{fig:5438_pv}%
\end{figure*}

\textbf{Disks:} \bba\ is able to reproduce the bulk of the emission in all spectral channels for the mock disks at all angular resolutions. Some residuals up to 8\,rms appear in the high-angular resolution data because of asymmetric emission in some spectral channels. On the contrary, the residuals are typically $\lesssim 4$\,rms at low-angular resolution because the asymmetries are smoothed out. In Fig. \ref{fig:5438_ch}, we show an example of data (black contours), model (red contours), and residuals for some representative spectral channels from the disk Petunia at high and low-angular resolution. For the same galaxy, in Fig. \ref{fig:5438_pv}, we show the emission in the major- and minor-axis PV diagrams at three different angular resolutions. We note that at high-angular resolutions, the PV along the major and minor axes have a typical shape of a rotating disk. As the angular resolution decreases, the emission in the PV diagrams becomes thicker. However, the symmetrical properties with respect to the axes defining the center and the systemic velocity (dashed lines) are conserved.

\begin{figure*}
        \centering
        \includegraphics[width=0.99\textwidth]{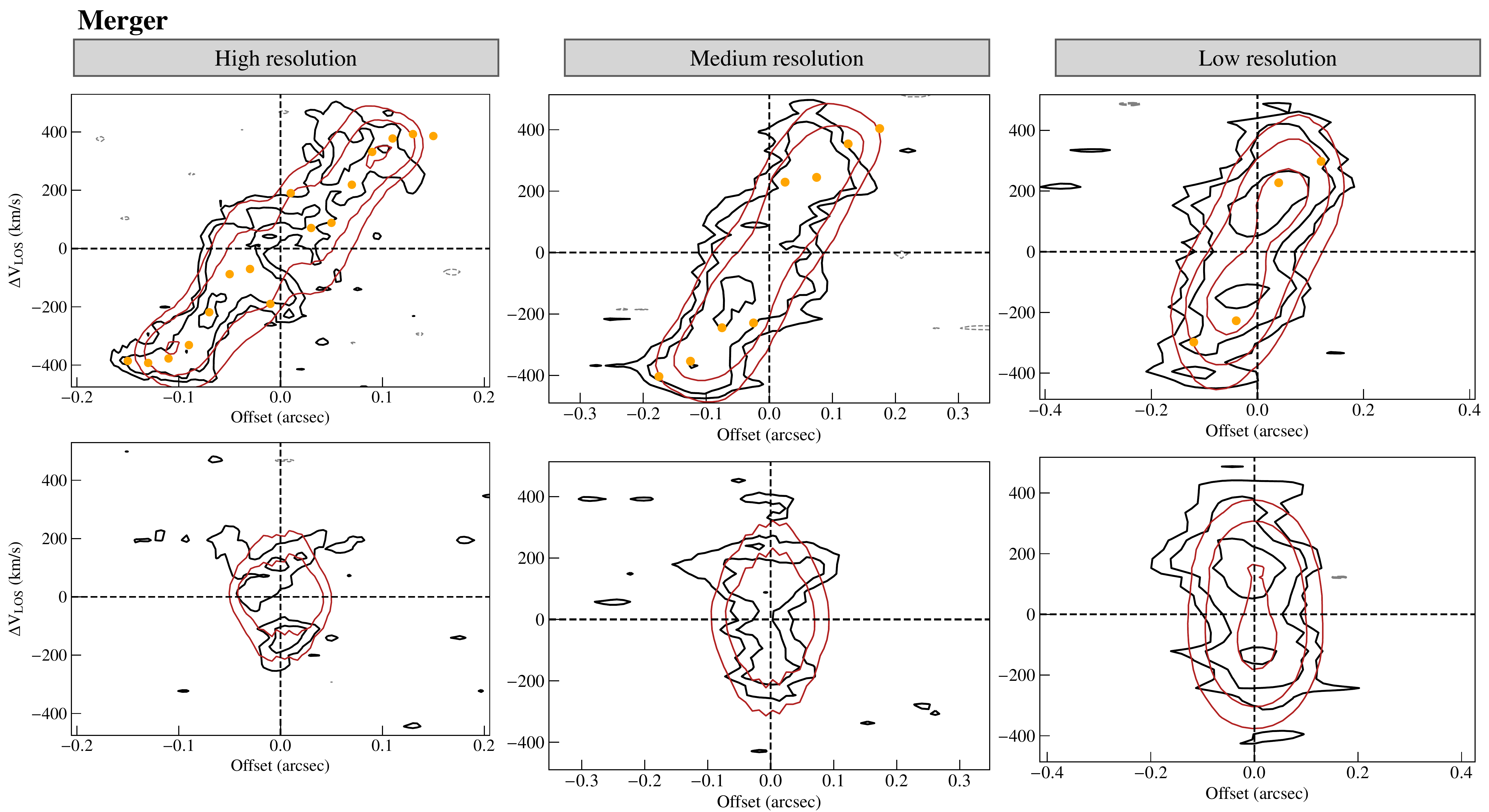}
        \caption{Same as in Fig. \ref{fig:5438_pv} but for the merging system Adenia at 60 $\deg$.}
        \label{fig:4294_pv}%
\end{figure*}

\begin{figure}
        \centering
        \includegraphics[width=0.49\textwidth]{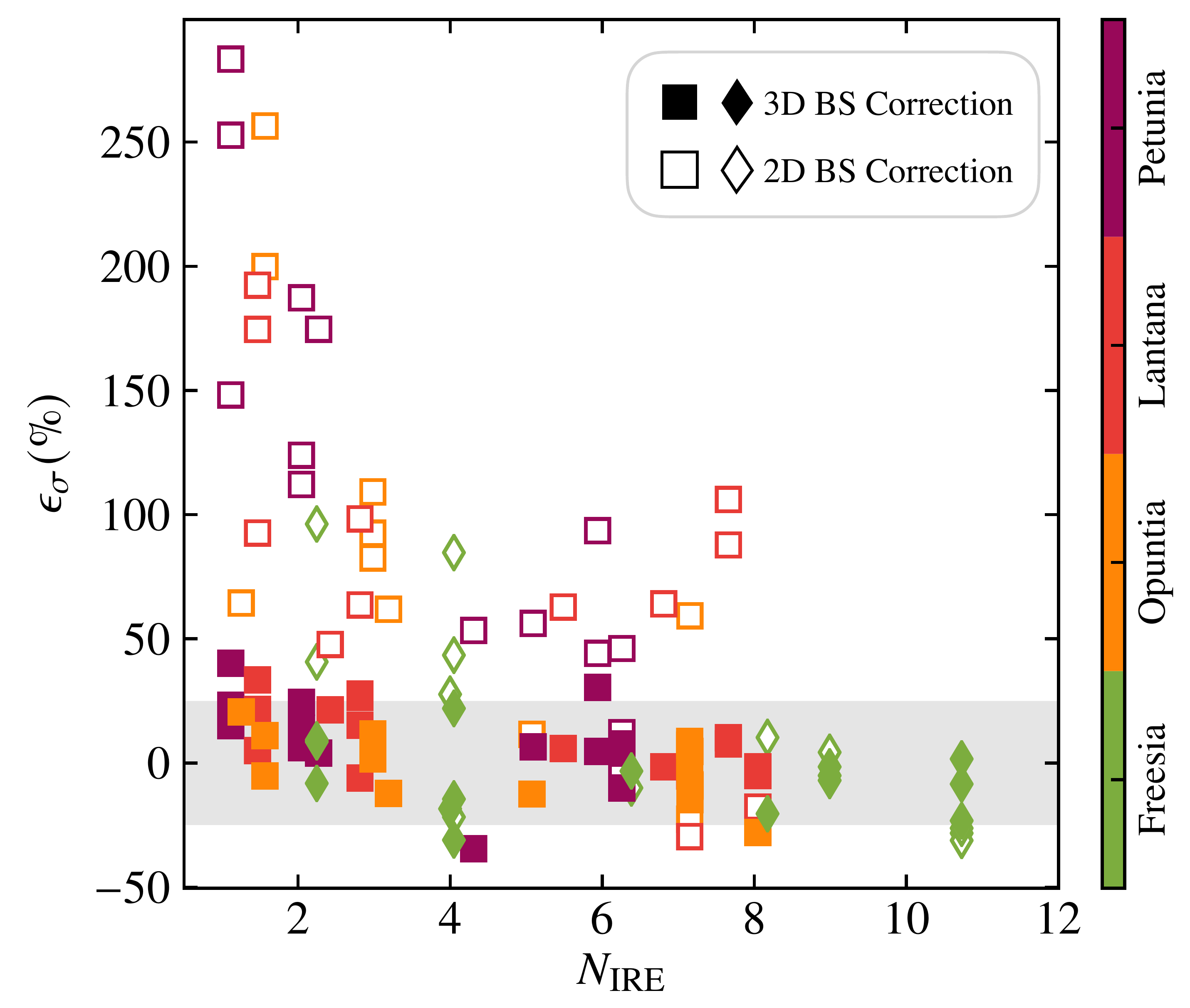}
        \caption{Relative errors of the recovered velocity dispersions, computed using eq. (\ref{eq:error}) for the disturbed disks Freesia (diamonds) and the three disks (squares) of our sample as a function of resolution (number of independent resolution elements along the semi-major axis).  Markers with the same color show the mock data for the same galaxy as indicated by the color-bar. The filled and empty markers show the errors for the velocity dispersions obtained with \bba\ (forward-modeling) and with the velocity-gradient beam-smearing correction. The gray areas mark the 25 percentage.}
        \label{fig:error}%
\end{figure}

To estimate the capability of \bba\ to recover reliable kinematic parameters for both high- and low-angular resolution observations, we computed the relative errors for the velocity dispersions:
    \begin{equation}
        \label{eq:error}
        \epsilon_{\sigma} = \frac{(\sigma_\mathrm{best}-\sigma_\mathrm{sim})}{\sigma_\mathrm{sim}}\,,
    \end{equation}
where $\sigma_\mathrm{best}$ is the radial average of the best-fit values derived with \bba\ and $\sigma_\mathrm{sim}$ is the average velocity dispersion from the moment-2 map of the face-on view\footnote{The selection of face-on view for determining the intrinsic value of velocity dispersion from the simulation is to minimize the contribution of bulk motions such as rotation to the observed $\sigma$.} of the simulated galaxy (noise-free and not convolved; see column 2 in Fig.2 of \citealt{Kohandel_2020} as an example).
In Fig. \ref{fig:error}, the filled markers show the values of $\epsilon_{\sigma}$ for all mock data created from the three disks and the disturbed disk Freesia as a function of the resolutions. The relative errors are within 25\% (gray area) in all cases, similar to the values reported in \citet{DiTeodoro_2015} for the tests on low-resolution observations of low-$z$ galaxies.
In the same Fig., the empty markers show the relative errors when $\sigma_\mathrm{best}$ is obtained by using a 2D technique for correcting a-posteriori for beam-smearing, the velocity gradient method (see Appendix \ref{appendix:gradient} for details). With this method, the velocity gradients across the beam in the moment-1 maps are subtracted linearly from the corresponding moment-2 maps, while no correction for beam-smearing is applied to the rotation velocity. The resulting average velocity dispersions are strongly biased to values larger than the intrinsic ones ($\sigma_\mathrm{sim}$), that is, the relative errors $\epsilon_{\sigma}$ reach values up to $\sim 200\%$ at low-angular resolutions.

\textbf{Mergers:} At high-angular resolution ($N_{\mathrm{IRE}} \gtrsim 4$), the chi-square values are a factor of $\gtrsim 10$ higher than those for disks at similar resolutions and S/N. The presence of two interacting systems, strong asymmetries, tidal features is evident in the data cubes, the PV diagrams, and the moment maps. Therefore, the disk model created by \bba\ is not able to reproduce the data, thus large residuals up to 40\,rms appear in some spectral channels (see bottom left panels in Fig. \ref{fig:5438_ch} as a representative example for the merger Adenia). As the resolution decreases, there are fewer residuals in all spectral channels (see bottom right panels in Fig. \ref{fig:5438_ch}), and the PV diagrams appear more regular and symmetric (right panels in Fig. \ref{fig:4294_pv}). However, the residuals of both the spectral channels and the PV diagrams are typically higher in the merger sample than the disk sample by a factor of $\approx 2$. This is further discussed and quantified in Sect. \ref{sec:new_par}.

\section{Kinematic classification from the literature}\label{sec:classification}

In this section, we summarize the most common methods that have been used in the literature to characterize the dynamical state of high-$z$ galaxies using emission-line observations. We then discuss the application of these methods to the mock data of \code{SERRA} galaxies and their ability to identify the fraction of disks in our sample correctly.
The classification schemes described in Sects. \ref{sec:k1} and \ref{sec:k2} are quantitative as their criteria are based on comparing kinematic measurements and physically- or empirically-defined thresholds. Such methods can be applied to the 84 mock data a) - d) described in Sect. \ref{sec:mock_data} and analyzed in Sect. \ref{sec:barolo}, having quality good enough for constraining a kinematic model.
On the contrary, the method described in Sect. \ref{sec:alpine} has been developed for classifying low-quality data: it is qualitative and mainly based on visual inspections of the kinematic maps. Accordingly, we adopt this method only on the 18 barely-resolved mock data, i.e., whose low angular resolution and S/N did not allow for recovering any kinematic properties (point e) in Sect. \ref{sec:mock_data}).

In Table \ref{tab:classification_overview}, we provide a summary of the data sets used for the analysis, the outcome, and recommendations on the data quality required to apply a given classification method. 

\begin{table*}
    \caption{Summary of the application of the classification techniques on our sample of mock data.\\
    $^\dagger$ a): Ideal data; b) ALMA high-angular resolution, high S/N data; c) ALMA medium- and low-angular resolution, high S/N; d) ALMA high- and medium-angular resolution, low S/N; e) ALMA barely-resolved, low S/N. For further details about the definition of the datasets, see Sect. \ref{sec:mock_data}. }
    \label{tab:classification_overview}
    \begin{center}
  \begin{tabularx}{\linewidth}{p{0.2\linewidth}|p{0.15\linewidth}|X}
    \hline
    Method & Testing dataset$^\dagger$ & Outcomes and Recommendations\\
    \hline
    ~ & ~ & \\
    Velocity fields and $V/\sigma$ (Sect. \ref{sec:k1}) & a), b), c), d) & 
    This method's ability to identify mergers strongly depends on the S/N and angular resolution of the data. 50\% of mergers in our sample are misclassified as disks at low angular resolutions. Disk galaxies are always classified correctly.  \\
    ~ & ~ & \\
    Kinemetry based (Sect. \ref{sec:k2}) & a), b), c), d) & 
    Up to 30\% of our mergers are misclassified as disks at medium- and low-angular resolutions. This method is optimal for performing a kinematic classification only when high-angular resolution ($N_{\mathrm{IRE}} \gtrsim 4$) data are available. The success rate of this method in distinguishing disks and mergers weakly depends on the S/N of the data. \\
    ~ & ~ & \\
    Qualitative (Sect. \ref{sec:alpine})& e) & 
    This method can result in a substantial underestimation of the disk fraction: $\approx$ 100\% of the disks in our sample are misclassified as mergers or dispersion-dominated galaxies. Instead, the fraction of mergers correctly classified is 50\%. Barely-resolved data with S/N $\lesssim 4$, typical of line-detection surveys, should not be employed to extract galaxies' dynamical states. \\
    ~ & ~ & \\
    \hline
     ~ & ~ & \\
    PVsplit (Sect. \ref{sec:newclas}) & a), b), c), d) & Observations of galaxies with at least 3 independent resolution elements along the major axis and S/N $\gtrsim 10$ guarantee the correct identification of disks and mergers. \\
    ~ & ~ & \\
    \hline
    \end{tabularx}
    \end{center}
\end{table*}

\subsection{Smooth velocity fields and $V/\sigma$ ratios}\label{sec:k1}

The most widely used method to classify high-$z$ star-forming galaxies is based on the visual inspection of the velocity maps and the computation of the $V/\sigma$ ratio \citep[e.g.,][]{Wisnioski_2015, Wisnioski_2018, Smit_2018}. Galaxies are classified as rotating disks if the following two criteria are satisfied: 1. moment-1 map with a smooth velocity gradient; 2. enough rotation support, i.e., $V/\sigma > 1.8$ \citep[e.g.,][]{Wisnioski_2015, Wisnioski_2018}. Galaxies with a smooth velocity map but with $V/\sigma < 1.8$ \citep[e.g.][]{Forster_2009, Law_2009, Kassin_2012, Jones_2021} are classified as dispersion-dominated systems, while galaxies with irregularities and no clear disk-like pattern in their velocity fields are mergers. These kinematic criteria have been used for both IFU and interferometric data and in a wide range of redshifts, from $z \sim 0.6$ up to $z \sim 7$ - 8 \citep[e.g.,][]{Forster_2009, Wisnioski_2015, Harrison_2017, Stott_2016, Smit_2018, Bakx_2020}. 
In addition to these criteria relying only on the information from the spatially-resolved kinematic maps, \citet{Wisnioski_2015} showed that if rest-frame spatially-resolved optical maps are available, two additional criteria based on the comparison between the geometrical properties of the optical and kinematic maps can help to constrain the fraction of disks and mergers. However, in this work, we investigate whether we can constrain the dynamical stage of galaxies when only emission line observations are available \citep{Jones_2021}.

\begin{figure}
   \centering
   \includegraphics[width=0.49\textwidth]{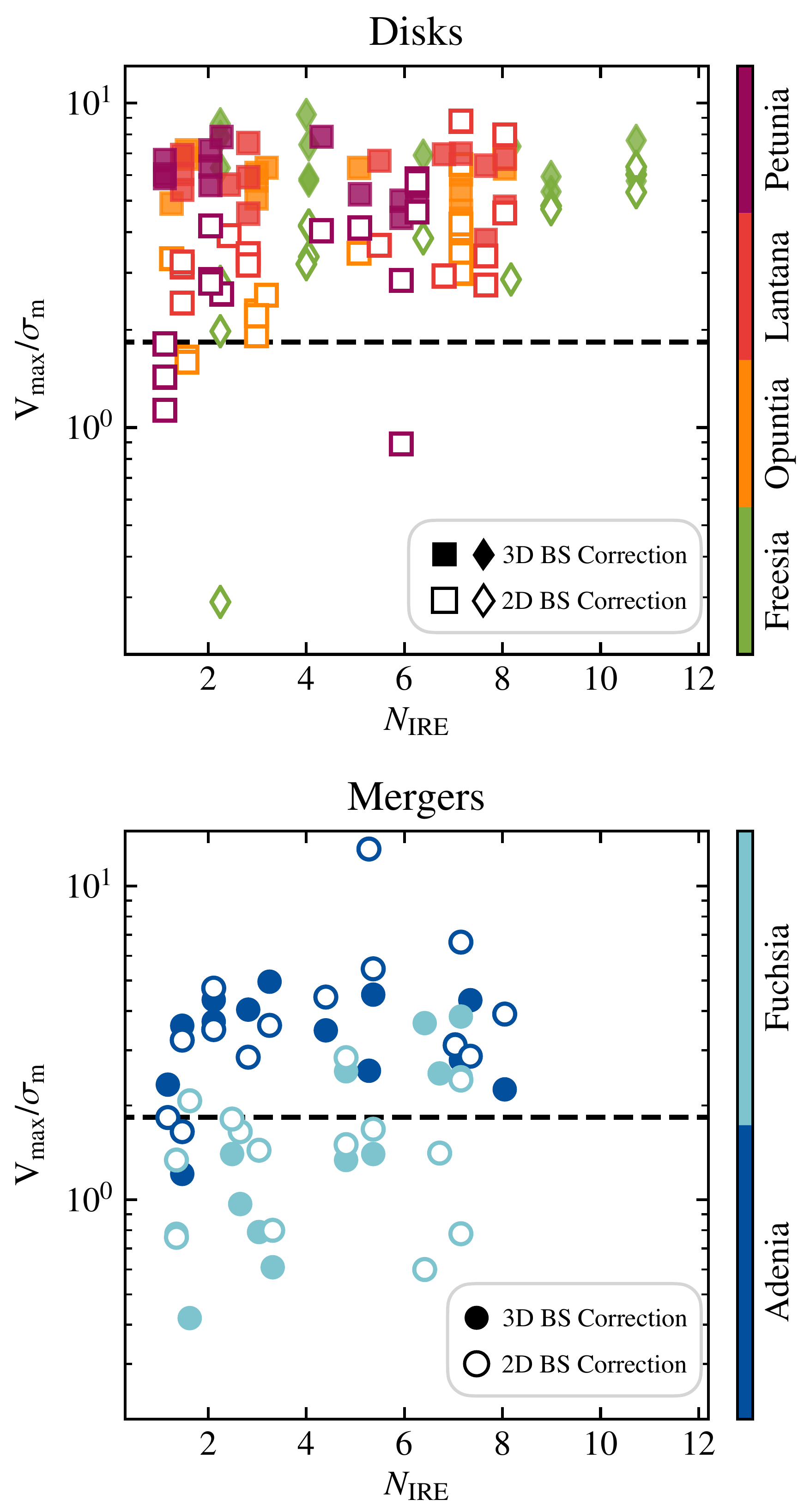}
   \caption{$V/\sigma$ ratio as a function of resolution (number of independent resolution elements along the semi-major axis) for the disks and the disturbed disk (upper panel) and the mergers (bottom panel). Markers with the same color show the mock data for the same \code{SERRA} galaxy as indicated by the color-bar.
   The black dashed line shows the threshold at 1.8. The filled markers show the $V/\sigma$ ratios obtained with \bba\, that is a 3D forward-modeling approach is applied to correct for the beam-smearing effect. The empty markers show the $V/\sigma$ obtained when a velocity-gradient method is employed to correct for the beam-smearing effect. \label{fig:vs}}
\end{figure}

To test whether this kinematic classification can distinguish the dynamical variety of our sample, we visually inspected the velocity fields of the mock data and computed the $V/\sigma$ ratios from the maximum value of the rotation velocity profiles and the median value of the velocity dispersion profiles. In Fig. \ref{fig:vs}, we show the values of $V/\sigma$ obtained both with \bba\ (filled markers) and by using the velocity gradient beam-smearing correction (empty markers) for the disks and mergers of our sample. The key results of this analysis are the following.

\begin{figure*}
   \centering
   \includegraphics[width=0.99\textwidth]{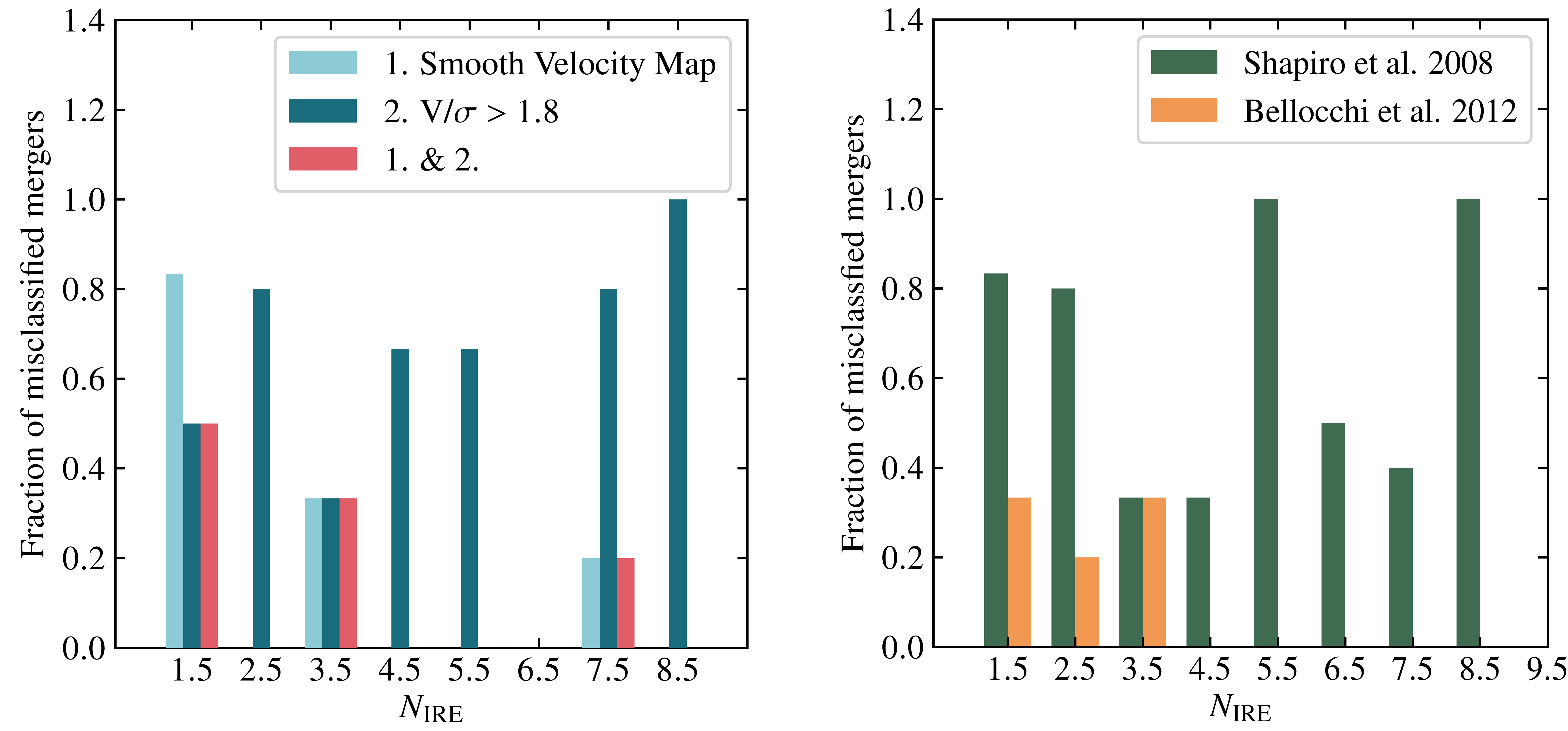}
   \caption{Fraction of mergers classified as disks based on the criteria described in Sects. \ref{sec:k1} (left, red bars) and \ref{sec:k2} (right) as a function of resolution. In the left panel, the light blue and dark green bars show the fraction of galaxies with smooth velocity maps and $V/\sigma \gtrsim 1.8$. The latter are obtained using the velocity-gradient method for correcting for beam smearing. The non-zero fraction at $N_{\mathrm{IRE}} \sim 7.5$ is due to the low S/N mock data. In the right panel, we show the fractions obtained using the \citet{Shapiro_2008} and \citet{Bellocchi_2012} methods as green and orange bars, respectively.
   \label{fig:fraction_k1_gradient}%
   }              
\end{figure*}

\begin{itemize}
\item The velocity fields of the mock disks and the disturbed disk Freesia have a smooth gradient and $V/\sigma$ ratios larger than 1.8 when the \bba\ best-fit parameters are used. In other words, all disks passed the disk criteria when $V$ and $\sigma$ are derived using the forward-modeling technique. However, the $V/\sigma$ ratios derived using the a-posteriori beam-smearing correction are systematically lower than those derived using \bba. In this case, at low resolution ($N_{\mathrm{IRE}} \lesssim$ 2), the fraction of disks with $V/\sigma > 1.8$ is 50\%, since the inferred velocity dispersions are significantly boosted and the inferred rotation velocities are underestimated due to the beam-smearing effect. Therefore, this combination leads to the misclassification of genuine disk galaxies as dispersion-dominated.
\item The velocity fields of the merger systems appear smooth or irregular depending on the data quality (angular resolution and S/N ratio), merger stage, and inclination. For example, at high-angular resolution ($N_{\mathrm{IRE}} \gtrsim 3$) the velocity field of the interacting system Adenia at $i\sim 60 \deg$ appears irregular or disk-like depending on the S/N ratios (Cols. 2 - 4 and 6 - 7 in Fig. \ref{fig:4294_moments}). However, the high S/N does not prevent the misclassification of Adenia as a disk at low-angular resolution at all inclinations (e.g., Col. 5 in \ref{fig:4294_moments}). The mock data for the merging system Fuchsia have, instead, a smooth gradient in the velocity field only at low-angular resolution ($N_{\mathrm{IRE}} \lesssim 2$) at $i \sim 60 \deg$ and $i \sim 80 \deg$. At $i \sim 30 \deg$, Fuchsia has an irregular velocity field at all S/N and angular resolutions. As a reference, the $V/\sigma$ ratios\footnote{Recall that $V$ and $\sigma$ for merger do not have a physical meaning since they are derived under the assumption that the fitting galaxy is a rotating disk (Sect. \ref{sec:mock_data}).} of the mergers cover a large range, from 0.4 to 5 and, contrary to the disks, there is no systematic difference between the values derived with the 2D and 3D beam-smearing correction methods.
\item In Fig. \ref{fig:fraction_k1_gradient} (left panel), we show the fraction of mergers passing the two criteria used for identifying disks, i.e., smooth velocity fields (cyan bars) and $V/\sigma > 1.8$ (blue bars). The red bars show the fraction of mergers that satisfy both the criteria and are hence misclassified as disks. At the typical angular resolutions of high-$z$ observations ($N_{\mathrm{IRE}} \lesssim 2$), the fraction of mergers misclassified as disks is 50\%.  The main reason for this behavior is that the moment maps are smoother and more regular as the resolution decreases. Our results are in agreement with the findings of \citet{Simons_2019}, who showed that at the typical angular resolutions of $z \sim 2$ IFU observations, merger galaxies pass the disk criteria with a probability from 10 to 100\% depending on the line of sight.
\end{itemize}

\subsection{Kinemetry-based methods}\label{sec:k2}
Developed by \citet{Krajnovic_2006}, the kinemetry method divides the moment-1 and moment-2 maps into concentric elliptical tilted rings from which azimuthal kinematic profiles are extracted and expanded into a finite number of harmonic terms
\begin{equation}
        \begin{split}
        K(R, \psi) & = A_0(R) + \sum_{i=1}^{N}A_{i}\cos(i\psi) + B_{i}\sin(i\psi) =\\ & = A_0(R) + \sum_{i=1}^{N}k_{i}\cos(i(\psi -\phi_{i}(R))\,,
        \end{split}
\end{equation}
where $k_{i} = \sqrt{A_i^2 + B_i^2}$ and $\phi_i = \arctan(A_i/B_i)$. For an ideal rotating disk, the only non-zero terms in the moment-1 maps are $A_0$, corresponding to the systemic velocity, and $B_1$, corresponding to the circular velocity. In the moment-2 maps, the only non-zero term is expected to be $A_0$, describing the variation of the velocity dispersion as a function of radius.
Therefore, asymmetries in the velocity and velocity dispersion maps are quantified by the presence of non-zero higher-order coefficients.

\citet{Shapiro_2008} proposed to use the average values $k_{i, avg, v}$(R) and $k_{i, avg, \sigma}$(R) of the higher-order coefficients\footnote{The higher-order coefficients $k_{i, 2, v}$ - $k_{i, 5, v}$ are used for the computation of $k_{i, avg, v} = (k_{i, 2, v}+k_{i, 3, v}+k_{i, 4, v}+k_{i, 5, v})/4$. $k_{i, avg, \sigma}$ is, instead, obtained by averaging the $k_{i, 1, \sigma}$ - $k_{i, 5, \sigma}$, $k_{i, avg, \sigma} = (k_{i, 1, \sigma}+k_{i, 2, \sigma}+k_{i, 3, \sigma}+k_{i, 4, \sigma}+k_{i, 5, \sigma})/5$}, to define the velocity and velocity dispersion asymmetry parameters,
\begin{subequations}\label{eq:v_and_sigma_shapiro}
\begin{equation}\label{eq:vasym_shapiro}
        v_{asym} = \frac{1}{N} \sum_{i=1}^{N}\left(\frac{k_{i, avg,v}}{B_{i, 1,v}}\right) 
\end{equation}
\begin{equation}\label{eq:sigmaasym_shapiro}
        \sigma_{asym} = \frac{1}{N} \sum_{i=1}^{N}\left(\frac{k_{i, avg,\sigma}}{B_{i, 1,v}}\right). 
\end{equation}
\end{subequations}
In Eqs. \ref{eq:v_and_sigma_shapiro}, the sum is over the number of rings $N$ into which the velocity and velocity dispersion maps are divided.
By using a series of template galaxies, cosmological simulations and actual observations of local disks and mergers, \citet{Shapiro_2008} defined an empirical threshold for the global asymmetric parameters, $K_{asym} = \sqrt{v_{asym}^2+\sigma_{asym}^2}$: galaxies with $K_{asym} > 0.5$ are classified as mergers, while galaxies with $K_{asym} < 0.5$ are disks. Using these criteria, the success rate in the correct classification of disks and mergers was estimated to be 89\% and 80\%, respectively \citep{Shapiro_2008}. 

\citet{Bellocchi_2012} proposed a modified computation of the asymmetry parameters, $v_{asym, w}$ and $\sigma_{asym, w}$, by giving more weight to asymmetric motions in the outer regions of galaxies. The global weighted asymmetric parameters, $K_{asym, w} = \sqrt{v_{asym,w }^2+\sigma_{asym, w}^2}$ should identify post-coalescence mergers, characterized by small asymmetries in the inner regions and outskirts with strong irregularities. \citet{Bellocchi_2012} adopted IFU observations from a sample of local ultra-luminous galaxies to define an empirical division of mergers and disks at $K_{asym, w} = 0.146$. However, they noticed that this value changes with the resolution: by redshifting their sample to $z = 3$, the value of $K_{asym, w}$ distinguishing the two kinematic classes is 0.13.

\begin{figure}[h!]
        \centering
        \includegraphics[width=0.49\textwidth]{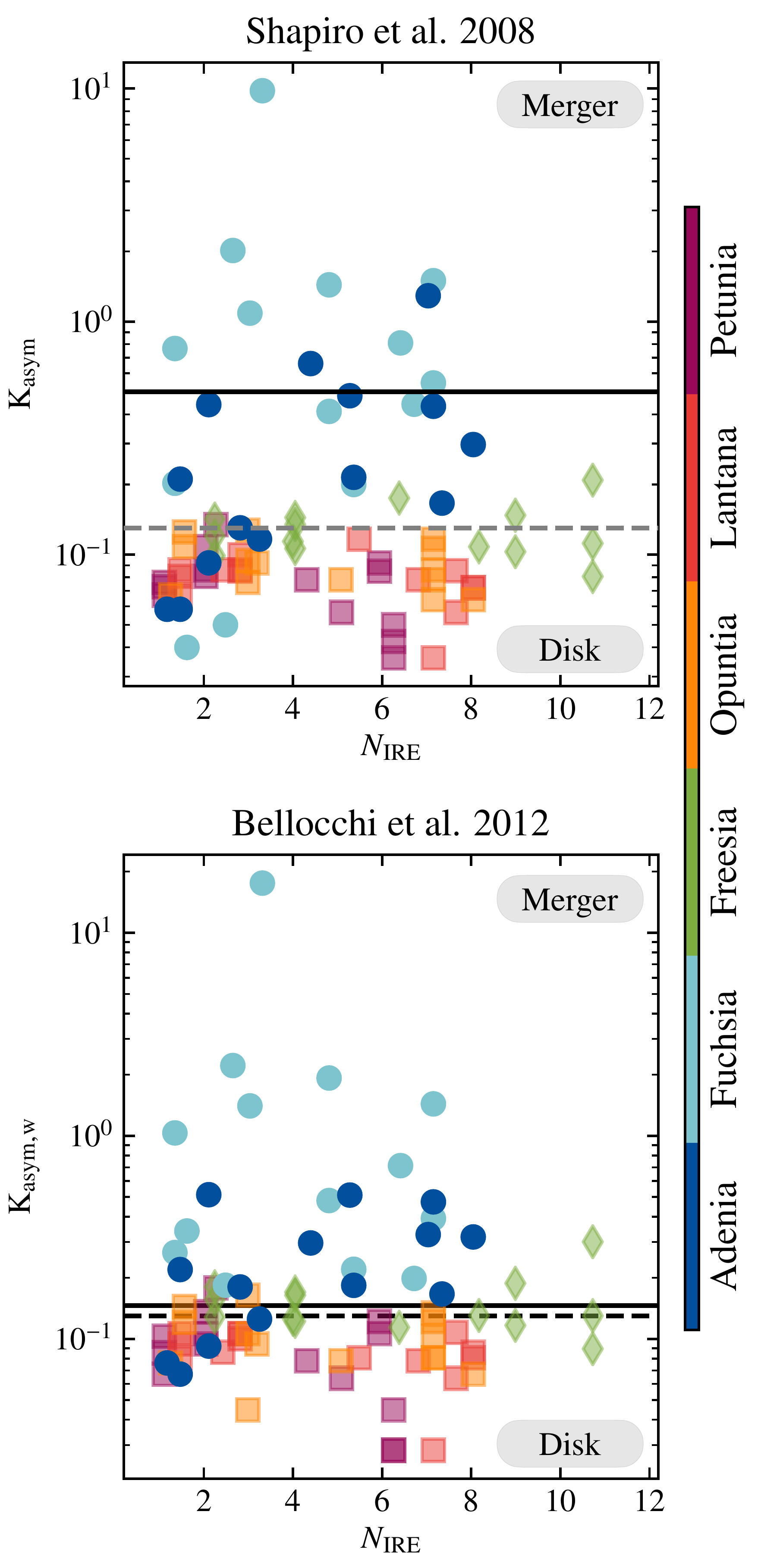}
        \caption{Global asymmetric parameters as a function of resolution for the disks (squares), mergers (circles), and the disturbed disk Freesia (diamonds). The global asymmetric parameters are computed according to the definitions of \citet{Shapiro_2008} (upper) and \citet{Bellocchi_2012} (bottom). The solid black line in the upper panel shows the limit of 0.5, indicated by \citet{Shapiro_2008} as the threshold for distinguishing between disks and mergers. In the bottom panel, the solid and dashed black lines show the limits of 0.146 and 0.13, indicated by \citet{Bellocchi_2012} as the discriminators between disks and mergers at low and high-$z$, respectively. As a reference, the limit of 0.13 is also shown in the upper panel (dashed gray line). 
        \label{fig:shapiro}
        }
\end{figure}

We can use our data sample to quantify the success rates in the kinematic classification for both \citet{Shapiro_2008} and \citet{Bellocchi_2012} methods. To define the rings used for the kinemetry decomposition, we assumed the best-fit center, inclination, and position angles found by {\bba}. The resulting $A_{i}$ and $B_{i}$ coefficients are then used to compute the asymmetry parameters and the values of $K_{asym}$ and $K_{asym, w}$. In Fig. \ref{fig:shapiro}, we show the global asymmetric parameters obtained using the definition from \citet{Shapiro_2008} and \citet{Bellocchi_2012} as a function of resolutions for our mock data. 
The key findings from the application of the kinemetry analysis are the following: 
\begin{itemize}
    \item 100\% of the disks are correctly classified by using both the criteria defined in \citet{Shapiro_2008} and \citet{Bellocchi_2012}. However, the values of $K_{asym, w}$ are consistent with the threshold of 0.13 for most galaxies at low angular resolution. On the contrary, the values of $K_{asym}$ are on average a factor of 5 lower than the threshold of 0.5.
    \item The same results are valid for the disturbed disk Freesia, except for the Ideal Mock at $i = 30\deg$, which features a value of $K_{asym, w} = 0.3$. For this line of sight and with the high sensitivity and angular resolution of the Ideal Mock, the \citet{Bellocchi_2012} method is able to identify the asymmetric motions in the faint outer regions of this galaxy.
    \item In Fig. \ref{fig:fraction_k1_gradient} (right panel), we show the fraction of mergers misclassified as disks according to the \citet{Shapiro_2008} and \citet{Bellocchi_2012} criteria. The latter is optimal for identifying the mergers of our sample at $N_{\mathrm{IRE}} \gtrsim 4$. On the contrary, the \citet{Shapiro_2008} method misclassifies mergers up to 100\% even for high-resolution observations. Note that the limit of $K_{asym} = 0.5$ from \citet{Shapiro_2008} was obtained after excluding the merger system IRAS 12112+0305 \citep{Garcia_2009} from their sample. \citet{Bellocchi_2012} noted that if the classification of IRAS 12112+0305 had been considered, the threshold $K_{asym}$ for the \citet{Shapiro_2008} sample would have been considerably lower. Interestingly, if we use the limit of 0.13 for $K_{asym}$, we obtain a fraction of misclassified mergers similar to that obtained by using the \citet{Bellocchi_2012} method. In other words, with the appropriate tuning of the threshold, the global asymmetric parameters $K_{asym}$ and $K_{asym, w}$ perform equally well in distinguishing between disks and mergers when high-angular resolution data are employed.
    \item At low angular resolution ($N_{\mathrm{IRE}} \lesssim 3.5$), the asymmetries in the velocity and velocity dispersion maps are washed out. The fraction of our mergers misclassified as disks range from 30 to 80\%, based on the \citet{Bellocchi_2012} and \citet{Shapiro_2008} methods, respectively. The latter is in agreement with the results of \citet{Goncalves_2010}, which reported a similar fraction of misclassified galaxies for a sample of $z \sim 2.2$ simulated galaxies. \citet{Shapiro_2008} and \citet{Wisnioski_2015} had already noticed that the most critical requirement in applying the kinemetry analysis is that the galaxy should be \quotes{well resolved}, with however no clear consensus on its definition. \citet{Shapiro_2008} defined a well-resolved galaxy to be covered by at least 3 resolution elements, while \citet{Wisnioski_2015} reported a value of 10 resolution elements. Our tests indicate that at least 8 independent resolution elements along the maximum extension of the galaxy (i.e., major-axis) are needed to identify the presence of interacting galaxies correctly.
\end{itemize}

\subsection{Qualitative method}\label{sec:alpine}

At z $\gtrsim 4$, a systematic kinematic classification of star-forming galaxies was performed using the ALPINE sample \citep{LeFevre_2020, Romano_2021, Jones_2021}. 
However, similar to most observations of galaxies at $z \gtrsim 4$ \citep[e.g.,][]{Smit_2018, Bakx_2020}, most of the ALPINE sample is barely resolved since the survey has been designed mainly for line detection.
As shown in \citet{Jones_2021}, only for part of the sample (14/75 galaxies), the S/N is high enough to adopt a quantitative kinematic classification (the one described in Sect. \ref{sec:k1}).

For the entire sample, \citet{Romano_2021} and \citet{LeFevre_2020} provide a qualitative classification by dividing galaxies into: mergers, disks, dispersion-dominated, and uncertain. In the classification scheme of \citet{Romano_2021} and \citet{LeFevre_2020}, mergers are galaxies showing irregular morphologies in the moment-0 maps, PV diagrams or optical images (when available, see discussion in Sec. \ref{sec:k1}); disturbed moment-1 maps; global line profile fitted by multiple components.
The disk class includes galaxies characterized by detection of the emission in contiguous channel maps; a gradient in the moment-1 maps; PV diagrams along the major axis with tilted emission; straight emission in the PV diagrams along the minor axis; possible double-horn profile; a single component in the optical images. Galaxies with no positional shifts of emission across channel maps; straight emission in the PV diagrams; a Gaussian line profile are assigned to the dispersion-dominated class.

\begin{figure*}
   \centering
   \includegraphics[width=0.99\textwidth]{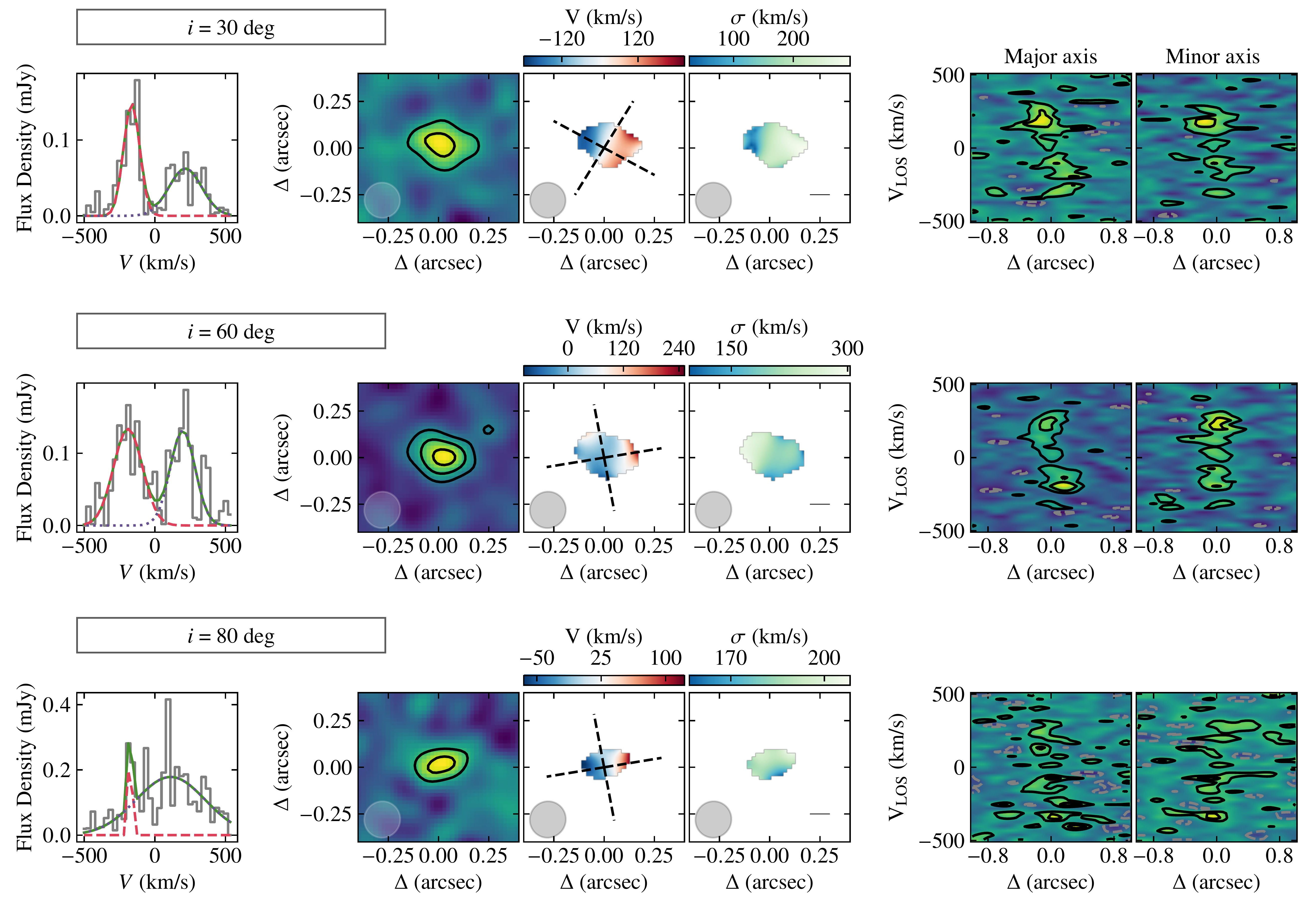}
   \caption{Kinematic maps and profile for the barely-resolved ALMA mock data for the disk Opuntia at 30 (upper panels), 60 (medium panels) and 80$\deg$ (bottom panels). Column 1: [CII] integrated spectrum (gray histogram) and best-fit model (solid green line), resulting from the sum of two components (red dashed and purple dotted lines). Column 2: moment-0 map with black contours showing the 3, 6 and 9\,rms levels. Columns 3 and 4: moment-1 and moment-2 maps obtained following the methodology in \citet{Romano_2021} (see Sect. \ref{sec:alpine} for details). The dashed lines in the moment-1 map show the directions of the major and minor axis along which the PV-diagrams are extracted. Columns 5 and 6: PV diagrams along the major (left) and minor (right) axis with contours at -2\,rms (dashed gray lines) and 2, 4, 8 \,rms (solid black lines).
   \label{fig:2208_qual}
   }
    \end{figure*}

\begin{figure*}
   \centering
   \includegraphics[width=0.99\textwidth]{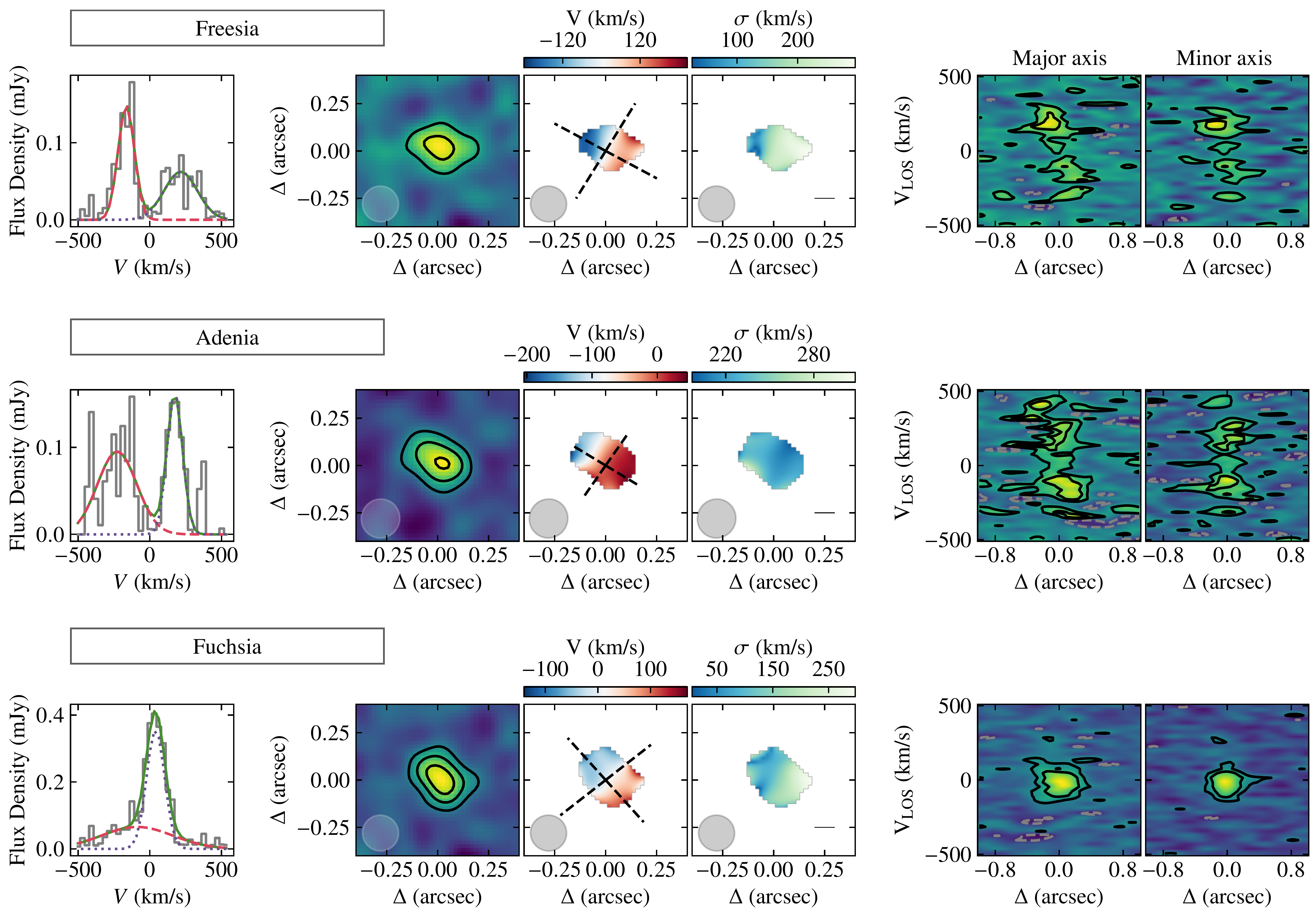}
   \caption{Same as in Figure \ref{fig:2208_qual}, but for the disturbed disk Freesia and the merging systems Adenia, Fuchsia at $i = 60\deg$.
    \label{fig:mergers_qual}
   }             %
\end{figure*}

Applying the kinematic criteria described in \citet{Romano_2021} and \citet{LeFevre_2020} requires a visual inspection of the moment maps, PV diagrams along the major and minor axis, and the integrated spectrum. Starting from the ALMA barely resolved mock cubes, we created the kinematic maps and profiles using the method described in \citet{Romano_2021} and summarized here:
\begin{itemize}
    \item the moment-1 and 2 maps are obtained after masking all pixels below 3\,rms in the moment-0 map.
    \item The moment-0 map is fitted using a two-dimensional Gaussian profile. The best-fit center is then used to extract the PV diagrams along the major and minor axis. The former is oriented along the direction of the strongest velocity gradient in the velocity map.
    \item The [CII] integrated spectrum is obtained after extracting the flux in the aperture defined by the 3\,rms-mask on the moment 0 map. One or multiple Gaussian components are used to fit the spectra.
\end{itemize}
In Fig. \ref{fig:2208_qual}, we show the [CII] integrated spectrum (gray line) and its best-fit model (solid green line), the moment maps, and the PV diagrams for one representative disk galaxy, the system Opuntia at three inclinations. Similar kinematic maps and profiles are shown in \ref{fig:mergers_qual} for the disturbed disk Freesia and the mergers Adenia and Fuchsia at $i = 60 \deg$.

The key findings resulting from applying the qualitative method are the following.
\begin{itemize}
    \item 8 out of 9 disks are classified as mergers, with the only exception of Opuntia at $i \sim 80\deg$ that is classified as a dispersion-dominated galaxy. At $i \sim 30$ and 60$\deg$, Opuntia is classified as a merger as two Gaussian components describe the integrated spectra, the velocity and velocity dispersion maps are disturbed, the PV diagrams along the major axis show two components at 3\,rms separated in velocity by at least 200 km s$^{-1}$ and with a spatial offset. At $i \sim 80\deg$, the integrated spectrum is, instead, well described by a single Gaussian line profile, and the PV diagrams are straight both along the major and minor axis. 
    \item The disturbed disk, Freesia, is classified as a disk at all inclinations.
    \item Adenia is correctly classified as a merger at all inclinations due to the multiple components in the integrated spectra and the PV diagrams.
    \item The merging system, Fuchsia, is classified as a dispersion-dominated galaxy at all inclinations due to the presence of a single component in the integrated spectra and straight PV diagrams.
\end{itemize}
To summarize, 100\% of the disks in our sample are misclassified as mergers or dispersion-dominated galaxies and only 50\% of the mergers are correctly identified. Even though an accurate statistical analysis is not feasible due to the low number of our mock data, the result advises against the adoption of the qualitative kinematic classification for low-quality data. This could be improved by considering the information from optical/UV maps, but it is out of the scope of the present work (see Sec. \ref{sec:k1}).

\section{PVsplit: a new 3D kinematic classification}\label{sec:newclas}

When a forward-modeling technique is adopted, robust kinematic parameters can be recovered even at low-angular resolutions and S/N $\gtrsim 5$ for disk galaxies (Sect. \ref{sec:kinematic_outputs}). 
However, with this data quality, the different techniques adopted so far failed in identifying the merging systems (Sect. \ref{sec:classification}) since the methods mainly use the information contained in the moment-1 and 2 maps. For example, one of the best classification schemes tested with our sample, the \citet{Bellocchi_2012} method, requires data at $N_{\mathrm{IRE}} \gtrsim 4$ to guarantee a robust characterization of the dynamical state of a galaxy.
With the sensitivities of the current instruments (e.g., ALMA), such high angular-resolution data with a minimum S/N of $\sim$ 5 require an integration time on the source of at least 30 hours for a typical Lyman-break galaxy at $z \sim 6$. Due to the typical observing time constraints for current programs, high-angular resolution observations can be obtained in the near future only by targeting a handful of Lyman-break galaxies at a time.

In this section, we present a proof-of-concept analysis aiming to show that a robust characterization of the dynamical state of a galaxy is still feasible with medium- and low-angular resolution data. The technique is called \quotes{PVsplit}, the main novelty is that it is based on the analysis of PV diagrams rather than using  moment-1 and 2 maps.
\subsection{PV parameters}\label{sec:new_par}

\begin{figure*}
   \centering
   \includegraphics[width=\textwidth]{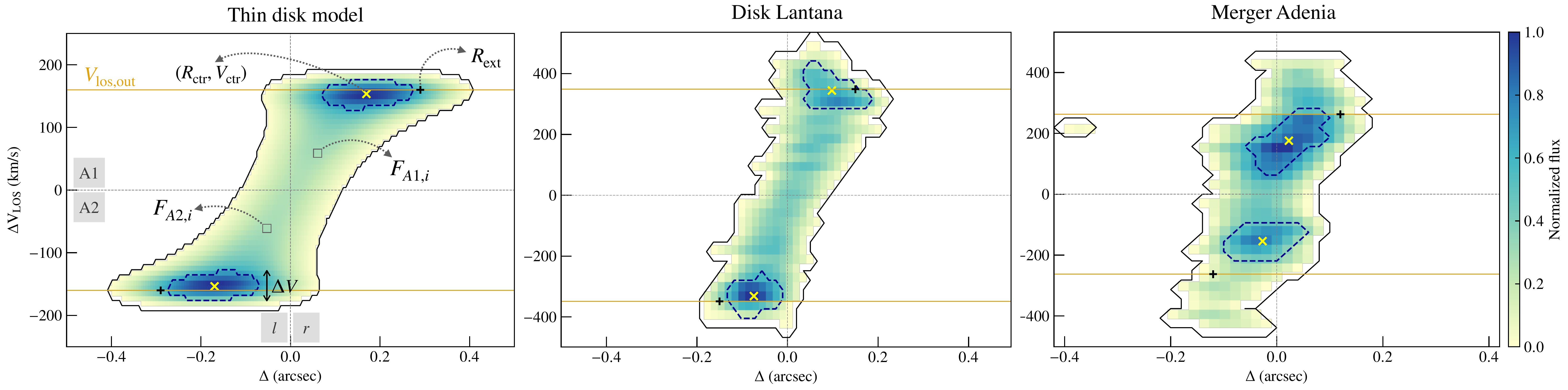}
   \caption{Schematic representation of the morphological and symmetrical features in the major axis PV diagrams considered for the computation of the $P$ parameters defined in Sect. \ref{sec:new_par}. The left panel shows the PV diagram for a representative thin rotating disk model along the major axis. The dotted gray horizontal line shows the position of the systemic velocity that divides the major-axis PV diagram into regions A1 and A2, while the dotted gray vertical lines indicate the center dividing A1 and A2 into two regions, $l$ and $r$. The $i$th representative pixels used for the computation of $P_{\mathrm{major}}$ are shown in the regions A1 and A2, respectively. The blue dotted contours enclose the 20\% brightest pixels, and the yellow crosses are their flux-weighted centroids. The black plus indicates the outermost radius used for the extraction of the kinematic parameters, and the solid orange lines are $V_{\mathrm{los, out}}$, the average values of the galaxy line-of-sight velocities. The double arrow indicates $\Delta V$, the extension along the velocity axis for one of the peak distributions. The major-axis PV diagrams from the high S/N, low-resolution mock data of the disk Lantana and merger Adenia are shown in the central and right panels.
   \label{fig:pvp}%
 } 
\end{figure*}

Based on the symmetry and morphology of the PV diagrams along the major and minor axes, we can define five new empirical parameters. A schematic representation of their main features is shown in Fig. \ref{fig:pvp} and a formal definition is given in the following:
\begin{itemize}
    \item $\mbox{\boldmath$P_{\mathrm{major}}$}$ quantifies the asymmetry of the major axis PV diagram with respect to the horizontal axis defining the systemic velocity. The latter divides the PV diagrams into two regions: A1 and A2 (see Fig. \ref{fig:pvp}). The parameter is defined as
    \begin{equation}
        P_{\mathrm{major}}=\log \left[ \frac{\sum_i (F_{A1,i}-F_{A2^*, i})^2}{\sum_{i} (F_{A1, i}+F_{A2^*, i})^2} \right]
    \end{equation}
    where $F_{A1,i}$ and $F_{A2^*, i}$ are the fluxes in the pixels $i$ of the region $A1$ and $A2^*$, respectively. The latter is the reflection of the region $A2$ with respect to the center. For an ideal rotating disk, $A2^* = A1$, thus $P_{\mathrm{major}}=0$.
  \item $\mbox{\boldmath$P_{\mathrm{minor, 1}}$}$ evaluates the asymmetries of the minor-axis PV diagram with respect to the systemic velocity.
  \item $\mbox{\boldmath$P_{\mathrm{minor, 2}}$}$ is similar to $P_{\mathrm{minor, 1}}$, but for quantifying the asymmetries of the minor-axis PV diagram with respect to the vertical axis defining the center.
  \item $\mbox{\boldmath$P_{\mathrm{V}}$}$ is specifically suited for medium and low-resolution observations.
  The emission peak in the PV diagrams has distinct morphologies in the disks and mergers both at high and low angular resolutions. The major-axis PV diagrams of the disks have a narrow emission peak along the velocity axis at values close to the line-of-sight velocities at the outermost radii of the galaxy, $V_{\mathrm{los, out}}$. Instead, for mergers, the peaks are i) spread over a large range of line-of-sight velocities, or ii) have a narrow distribution closer to the systemic velocity than to $V_{\mathrm{los, out}}$. For quantifying these different distributions of the emission peaks along the velocity axis of the PV diagrams, we first selected the 20\% brightest pixels in the regions A1 and A2 separately.
  These pixels are used to compute $P_{\mathrm{V}}$ as
  \begin{equation}
  \label{eq:pvpv}
      P_{\mathrm{V}} =\log \left| \mathrm{Mean_{A1,A2}} \left[ \frac{V_{\mathrm{ctr}}}{V_{\mathrm{los, out}}} \exp{\left(-\frac{\Delta V}{V_{\mathrm{ctr}}}\right)} \right] -1 \right|,
  \end{equation}
  where $V_{\mathrm{ctr}}$ is the flux-weighted average line-of-sight velocity, $\Delta V$ is the difference between the 97.7\% and 2.3\% of the peak distributions along the velocity axis. The mean in Eq. (\ref{eq:pvpv}) is over the two values of $P_{\mathrm{V}}$ obtained for the regions A1 and A2. The maximum value is considered when multiple, distinct emission peaks are identified in A1 and A2. The exponential term is close to 1 in the case of a narrow emission peak distribution, typical of a disk, and it is $\gg 1$ for a peak emission spread over a large range of velocities, typical of a merger.
  \item $\mbox{\boldmath$P_{\mathrm{R}}$}$ is similar to $P_{\mathrm{V}}$ but for the distribution of the emission peak along the radial axis. To describe this empirical parameter, we define the two quadrants $l$ and $r$ over which the vertical line defining the center of the major-axis PV diagrams splits the regions A1 and A2 into two parts. The emission peak covers either quadrant $l$ or $r$ for the disks. Instead, for the mergers, the emission peak covers a region that is spread over both the $l$ and $r$ quadrants, or it is closer to the center than to the outermost radius used for the extraction of the kinematic parameters, $R_{\mathrm{ext}}$. To define $P_{\mathrm{R}}$, we used the following equation:
    \begin{equation}
       P_{\mathrm{R}} = \mathrm{Mean_{A1,A2}} \left( \frac{R_{\mathrm{ctr}}}{R_{\mathrm{ext}}}\right),
    \end{equation}
   where $R_\mathrm{ctr}$ is the flux-weighted average radial position of 20\% brightest pixels in the regions A1 and A2.
\end{itemize}
To measure the PV parameters for our mock data, we use the PV diagrams extracted along the best-fit major and minor axes found with \bba. Since, as discussed in Sect. \ref{sec:barolo}, the kinematic fitting is not feasible for the barely-resolved mock data (point e) in Sect. \ref{sec:mock_data}), they are excluded from the analysis and discussion in this Section.

\subsection{Construction of the classifier}\label{sec:stat_new}

\begin{figure}
   \centering
   \includegraphics[width=0.5\textwidth]{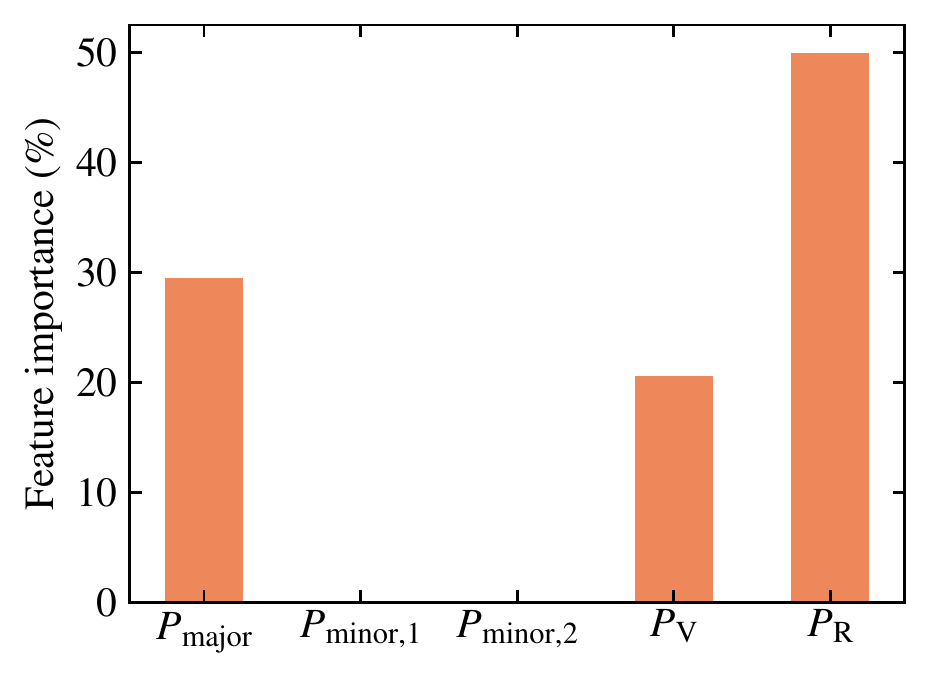}
   \caption{Importance of the different PV parameters for the disk/merger classification. The fractional importance is computed for the different parameters (definitions in Sec. \ref{sec:new_par}) by constructing a decision tree classifier to distinguish between disks and mergers.
   The decision tree uses our sample of 84 galaxies with different resolutions and S/N, using $70\%$ ($30\%$) of the data for training (testing), resulting in a $R^2$ score of 84\%.
   \label{fig:feature_importance}
   }
\end{figure}

\begin{figure*}
   \centering
    \includegraphics[scale=0.7]{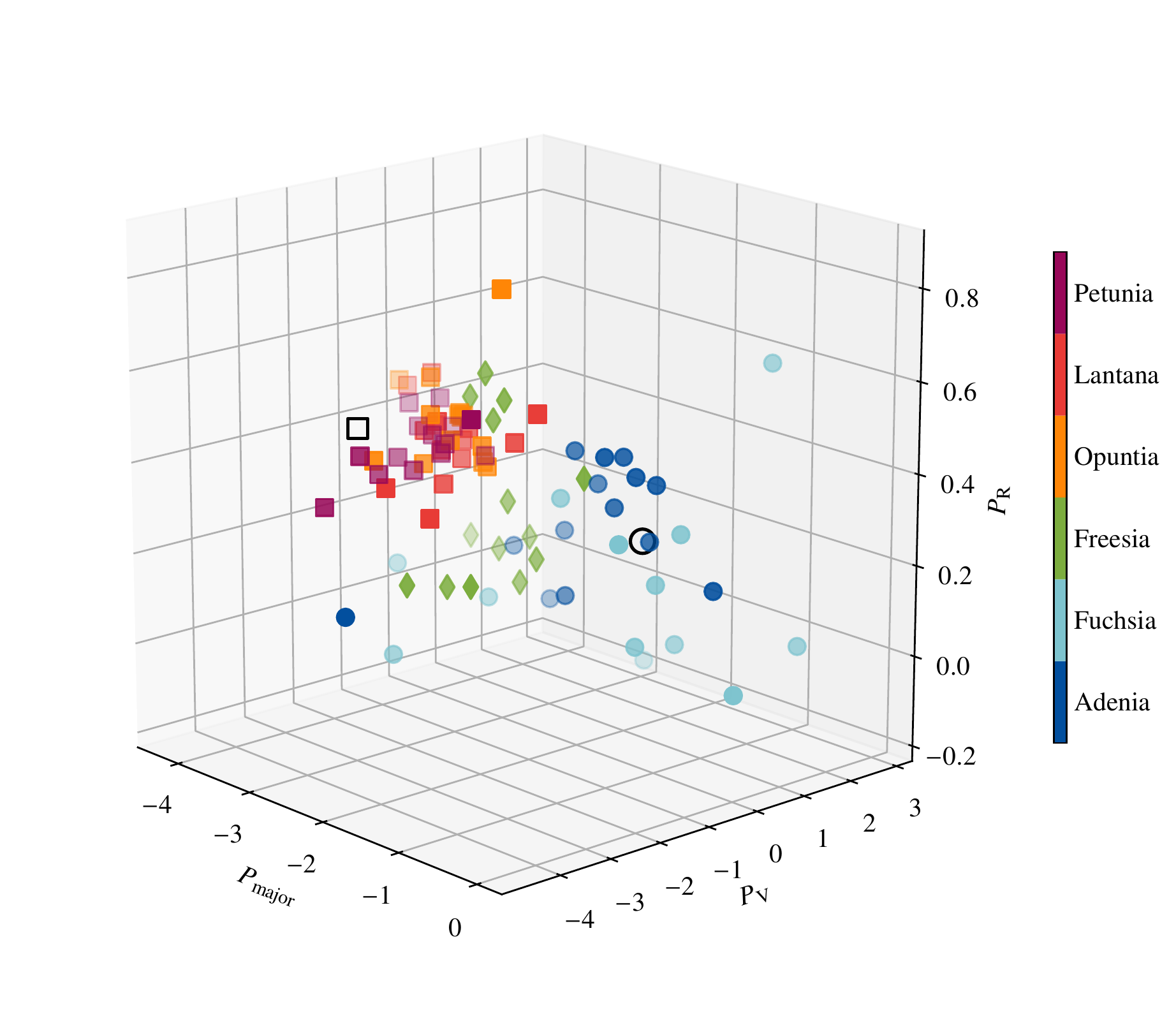}
    \caption{Position of the 84 mock data (squares: disks, circles: mergers; diamonds: disturbed disk Freesia) in the PVsplit parameter space. The empty black square and circle show the position of a spiral and a merger galaxy at $z = 0$, respectively.
    \label{fig:pvcube}%
    }
\end{figure*}

The PV parameters defined in Sect. \ref{sec:new_par} have different trends as a function of the resolution: this complex situation can be appreciated in Fig. \ref{fig:pvN}.
Visual inspections indicate that $P_{\mathrm{minor, 1}}$ and $P_{\mathrm{minor, 2}}$ -- the asymmetric parameters along the minor axis -- have overlapping values for disks and mergers, regardless of the resolution. Instead, the disk and merger populations seems to be located in different regions for the parameters $P_{\mathrm{major}}$, $P_{\mathrm{V}}$, and $P_{\mathrm{R}}$.
In particular, $P_{\mathrm{major}}$ seems able to clearly distinguish between disks and mergers when $N_{\mathrm{IRE}} \gtrsim 3$; however, for low resolution ($N_{\mathrm{IRE}} \lesssim 3$), both disks and mergers have values of $P_{\mathrm{major}} \lesssim 0.25$, that is, the major-axis PV diagrams of the mergers are as symmetric as PV diagrams for disks.

The insight given by this visual inspection can be formalized by constructing a machine learning (ML) classifier.
We build a classifier based on a decision tree using the \code{scikit-learn} implementation \citep{pedregosa:2012}. In general, a decision tree is a data-driven (supervised) ML algorithm: given a set of parameters (features in the ML language), and a discrete number of classes (labels in ML), the algorithm splits the parameter space into sub-volumes with maximal information content; the final goal of the algorithm is to predict the class with arbitrary parameters, i.e., that have not been used in the construction of the tree.
Our classes are merger and disk, and we select the 5 PV parameters described in Sect. \ref{sec:new_par} and use the 84 galaxies as our data set. By dividing the sample randomly, 70\% of the data is used as training in the construction of the algorithm, 30\% as a testing set.

The resulting decision tree has a $R^2$ score of 84\%, i.e., it is reasonably reliable in distinguishing between disk and mergers, given the small data set we are adopting here.
We then checked the feature importance of our parameters in distinguishing if a galaxy is a disk or a merger. In Fig. \ref{fig:feature_importance} we plot the importance of each PV parameter used in the ML. Both $P_{\mathrm{minor, 1}}$ and $P_{\mathrm{minor, 2}}$ appear to have negligible importance, while $P_{\mathrm{major}}$, $P_{\mathrm{R}}$ and  $P_{\mathrm{V}}$ are dominant in determining the kinematic class of the galaxy.
These results confirm our visual intuition. However, we have to extend the size of the data set to build a more reliable classifier (e.g., $R^2\gtrsim 95\%$) and eventually explore the importance of additional parameters in the PV space; this is left for future work.

In Fig. \ref{fig:pvcube}, we show the position of the 84 mock \code{SERRA} galaxies in the 3D parameter space, defined by the three parameters with higher importance, i.e. $P_{\mathrm{major}}$, $P_{\mathrm{V}}$, $P_{\mathrm{R}}$..
The merger (Adenia, Fuchsia) and disk systems (Opuntia, Lantana, Petunia) are clearly clustered in two different regions. The disturbed disk Freesia (diamonds) data is spread over the disk and merger regions: at low resolution, the galaxy behaves like a disk, while at high resolution, the galaxy is classified as a merger, depending on the inclination.
As a validation, we computed the PV parameters from real observations of a prototypical disk and a merger at $z = 0$ (see Appendix \ref{appendix:lowz} for details):
\begin{itemize}
    \item the spiral galaxy, NGC\,2403, for which we used HI observations at low-angular resolution \citep[empty black square,][]{DeBlok_2014}.
    \item The Antennae galaxy, a prototypical example of a merger system in the local Universe, as observed at low-angular resolution. ALMA Science Verification observations of the CO(3-2) emission line are employed for creating these data\footnote{\href{https://almascience.nrao.edu/alma-data/science-verification/antennae-galaxies}{https://almascience.nrao.edu/alma-data/science-verification/antennae-galaxies}}. 
\end{itemize}
The low-$z$ spiral and the Antennae galaxy are in the disk and merger regions defined by the \code{SERRA} disks and merger distributions, respectively (empty markers in Fig. \ref{fig:pvcube}). This analysis shows that by using PVsplit on data with S/N$\gtrsim 10$ and with at least 3 independent resolution elements along the major axis, one can accurately characterize the dynamical stage of a galaxy in a way that is independent of the redshift and tracers employed to measure the kinematics. A summary of the data sets used for the analysis of PVsplit, the main results, and recommendations on the required data quality is shown in Table \ref{tab:classification_overview}. 

\subsubsection{Advantages of PVsplit}

As discussed in Sect. \ref{sec:classification}, the common feature of the classification methods mostly used until now is the analysis of moment maps, which are projections of the data cubes to 2D space obtained after masking low S/N pixels in each spectral channel. In these 2D maps, the possibility to constrain the presence of non-circular motions driven by interactions strongly depends on the quality of the data. At medium and low-angular resolution, most asymmetries in the moment maps are smoothed out so that identifying a merging system becomes challenging.

The main novelty of PVsplit with respect to the standard classification methods is that it leverages the constraints on the dynamical stages of galaxies contained in the cubes, the native space of the data, through the analysis of the PV diagrams. The latter can be considered a more accurate representation of the data cubes since they require only to define the position angle along which they are extracted. In addition to the symmetrical properties of the PV diagrams along the major axis, PVsplit relies on the morphological features of the emission peaks. The latter is especially crucial at the low-angular resolution, showing that the non-circular motions driven by interactions are still imprinted in the data.

\section{Summary}\label{sec:conclusions}

Characterizing the dynamical state of galaxies is one of the critical steps to constrain their mass assembly history. This is particularly crucial in the early Universe, where the build-up activity was considerably accelerated with respect to the present-day rate. We are now entering a new era for this field since thousands of galaxies up to $z \sim 7$ will be discovered with the current and future facilities \citep[e.g., JWST, WFIRST, Euclid,][]{Mason_2015, Cuby_2019, Vogelsberger_2020}. These datasets will be a pool for finding the ideal targets for spatially-resolved follow-up observations. Defining an observational strategy to constrain the growth of these first galaxies is critical to fully taking advantage of this opportunity.

In this paper, we have analyzed the performance of three methods commonly used to constrain the dynamical properties of intermediate- and high-$z$ galaxies depending on angular resolution and S/N of the data. We have used six $z \sim$ 6 - 7 Lyman Break galaxies at three viewing angles from the \code{SERRA} zoom-in cosmological simulation suite \citep{Pallottini_2022}. The selected \code{SERRA} galaxies cover a range of dynamical states: three are isolated disks, one is a disturbed disk, and two are major mergers. We created mock observations of the [CII] emission with various data quality \citep{Kohandel_2020}: from an ideal case with very high S/N and angular resolutions to ALMA mock data typical of line-detection surveys. The key findings are:
\begin{itemize}
    \item The kinematic classification scheme proposed by \citet{Romano_2021} for low-SN and barely-resolved data (i.e., for a given galaxy, the area at S/N$\gtrsim 3$ extends for less than 50\% of the beam area in each spectral channel) may significantly underestimate the fraction of disk galaxies. We discourage any kinematic analysis on such observations since the resulting dynamical stages can be strongly biased and difficult to control, even when large statistics are available.
    \item The classification scheme (smooth velocity field and $V/\sigma$ ratio) proposed by \citet{Wisnioski_2015} and \citet{Wisnioski_2018} underestimates the fraction of disks when the values of $V$ and $\sigma$ are obtained using a 2D beam-smearing correction. On the contrary, the disks are correctly identified when a 3D forward-modeling technique is employed. The success rate of this method in identifying merging systems strongly depends on the S/N, even when high angular resolutions are used.  
    \item The kinemetry-based classification methods \citep{Shapiro_2008, Bellocchi_2012} underestimate the fraction of mergers at the typical angular resolution of high-$z$ observations. However, the \citet{Bellocchi_2012} method is optimal for high-S/N and high angular resolution data (i.e., S/N $\gtrsim 5$ and at least 4 independent resolution elements along the semi-major axis of the galaxy), that is, with data quality currently available only for a handful of targets with the current facilities (e.g., ALMA). 

\item Our analysis showed that the classical classification methods are sub-optimal to correctly characterize the dynamical stage as they analyzed the moment maps. This paper showed how a change of perspective is sufficient to overcome the limitations of the standard kinematic classification methods. Instead of the moment maps, the PV diagrams contain a wealth of information that allows accurate characterization of the kinematic state of a galaxy even at low-angular resolution. To quantify this, we have introduced here a new classification method, PVsplit, that relies on three empirical parameters describing the asymmetry and morphological features of the PV diagram along the galaxy's major axis. PVsplit requires a minimum of $\approx 3$ independent resolution elements along the major axis of a galaxy and S/N, computed as the the median of the maximum S/N in each
spectral channel, of at least 10 to correctly identify its dynamical stage as a disk or a merger. With ALMA, such data quality can be achieved with an integration time of 7 hours for a $z \sim 6$ main-sequence galaxy with SFR $\simeq 30\, \msolar \rm{yr}^{-1}$ and $\rm{L}_{\rm{[CII]}} \simeq 3 \times 10^{8}\rm{L}_{\odot}$. In addition, PVsplit is applicable to galaxies at all redshifts and with kinematics derived from gas emission lines. 
\end{itemize}

In the second paper of this series, we will expand our results of PVsplit on a large sample of simulated and observed galaxies with the aim of statistically evaluating the performance of this new method using a more extensive range of dynamical states. In that work, we will also discuss the position of observed low- and high-$z$ galaxies in the PVsplit parameter space.


\begin{acknowledgements}
FR thanks F. Fraternali for providing the HI data for NGC2403 and D. Krajnovi{\'c} for providing the python version of \code{KINEMETRY}. FR is grateful to F. Fraternali and M. Romano for useful comments and discussions.
FR acknowledges support from the European Union’s Horizon 2020 research and innovation program under the Marie Sklodowska-Curie grant agreement No. 847523 ‘INTERACTIONS’ and the Cosmic Dawn Center that is funded by the Danish National Research Foundation under grant No. 140.\\
MK, AP, AF, LV acknowledge support from the ERC Advanced Grant INTERSTELLAR H2020/740120 (PI: Ferrara). Any dissemination of results must indicate that it reflects only the author’s view and that the Commission is not responsible for any use that may be made of the information it contains.
We acknowledge the CINECA award under the ISCRA initiative, for the availability of high performance computing resources and support from the Class B project SERRA HP10BPUZ8F (PI: Pallottini).
We gratefully acknowledge computational resources of the Center for High Performance Computing (CHPC) at SNS.\\
This paper makes use of the following ALMA data: ADS/JAO.ALMA\#2011.0.00003.SV. ALMA is a partnership of ESO (representing its member states), NSF (USA) and NINS (Japan), together with NRC (Canada), NSC and ASIAA (Taiwan), and KASI (Republic of Korea), in cooperation with the Republic of Chile. The Joint ALMA Observatory is operated by ESO, AUI/NRAO and NAOJ.
We acknowledge usage of the Python programming language \citep{python2,python3}, Astropy \citep{astropy}, Cython \citep{cython}, Matplotlib \citep{matplotlib}, NumPy \citep{numpy}, \code{pynbody} \citep{pynbody}, \code{scikit-learn} \citep{pedregosa:2012}, and SciPy \citep{scipy}.

\end{acknowledgements}


\bibliographystyle{aa}
\bibliography{ms.bib} 

\begin{thebibliography}{101}
\expandafter\ifx\csname natexlab\endcsname\relax\def\natexlab#1{#1}\fi

\bibitem[{{Agertz} {et~al.}(2013){Agertz}, {Kravtsov}, {Leitner}, \&
  {Gnedin}}]{Agertz_2013}
{Agertz}, O., {Kravtsov}, A.~V., {Leitner}, S.~N., \& {Gnedin}, N.~Y. 2013,
  \apj, 770, 25

\bibitem[{{Asplund} {et~al.}(2009){Asplund}, {Grevesse}, {Sauval}, \&
  {Scott}}]{Asplund_2009}
{Asplund}, M., {Grevesse}, N., {Sauval}, A.~J., \& {Scott}, P. 2009, \araa, 47,
  481

\bibitem[{{Astropy Collaboration} {et~al.}(2013){Astropy Collaboration},
  {Robitaille}, {Tollerud}, {Greenfield}, {Droettboom}, {Bray}, {Aldcroft},
  {Davis}, {Ginsburg}, {Price-Whelan}, {Kerzendorf}, {Conley}, {Crighton},
  {Barbary}, {Muna}, {Ferguson}, {Grollier}, {Parikh}, {Nair}, {Unther},
  {Deil}, {Woillez}, {Conseil}, {Kramer}, {Turner}, {Singer}, {Fox}, {Weaver},
  {Zabalza}, {Edwards}, {Azalee Bostroem}, {Burke}, {Casey}, {Crawford},
  {Dencheva}, {Ely}, {Jenness}, {Labrie}, {Lim}, {Pierfederici}, {Pontzen},
  {Ptak}, {Refsdal}, {Servillat}, \& {Streicher}}]{astropy}
{Astropy Collaboration}, {Robitaille}, T.~P., {Tollerud}, E.~J., {et~al.} 2013,
  \aap, 558, A33

\bibitem[{{Bakx} {et~al.}(2020){Bakx}, {Tamura}, {Hashimoto}, {Inoue}, {Lee},
  {Mawatari}, {Ota}, {Umehata}, {Zackrisson}, {Hatsukade}, {Kohno}, {Matsuda},
  {Matsuo}, {Okamoto}, {Shibuya}, {Shimizu}, {Taniguchi}, \&
  {Yoshida}}]{Bakx_2020}
{Bakx}, T. J.~L.~C., {Tamura}, Y., {Hashimoto}, T., {et~al.} 2020, \mnras, 493,
  4294

\bibitem[{Behnel {et~al.}(2011)Behnel, Bradshaw, Citro, Dalcin, Seljebotn, \&
  Smith}]{cython}
Behnel, S., Bradshaw, R., Citro, C., {et~al.} 2011, Computing in Science
  Engineering, 13, 31

\bibitem[{{Bellocchi} {et~al.}(2012){Bellocchi}, {Arribas}, \&
  {Colina}}]{Bellocchi_2012}
{Bellocchi}, E., {Arribas}, S., \& {Colina}, L. 2012, \aap, 542, A54

\bibitem[{{B{\'e}thermin} {et~al.}(2020){B{\'e}thermin}, {Fudamoto}, {Ginolfi},
  {Loiacono}, {Khusanova}, {Capak}, {Cassata}, {Faisst}, {Le F{\`e}vre},
  {Schaerer}, {Silverman}, {Yan}, {Amorin}, {Bardelli}, {Boquien}, {Cimatti},
  {Davidzon}, {Dessauges-Zavadsky}, {Fujimoto}, {Gruppioni}, {Hathi}, {Ibar},
  {Jones}, {Koekemoer}, {Lagache}, {Lemaux}, {Moreau}, {Oesch}, {Pozzi},
  {Riechers}, {Talia}, {Toft}, {Vallini}, {Vergani}, {Zamorani}, \&
  {Zucca}}]{Bethermin_2020}
{B{\'e}thermin}, M., {Fudamoto}, Y., {Ginolfi}, M., {et~al.} 2020, \aap, 643,
  A2

\bibitem[{{Bosma}(1978)}]{Bosma_1978}
{Bosma}, A. 1978, PhD thesis, -

\bibitem[{{Bouch{\'e}} {et~al.}(2015){Bouch{\'e}}, {Carfantan}, {Schroetter},
  {Michel-Dansac}, \& {Contini}}]{Bouche_2015}
{Bouch{\'e}}, N., {Carfantan}, H., {Schroetter}, I., {Michel-Dansac}, L., \&
  {Contini}, T. 2015, \aj, 150, 92

\bibitem[{{Bournaud} {et~al.}(2007){Bournaud}, {Elmegreen}, \&
  {Elmegreen}}]{Bournaud_2007}
{Bournaud}, F., {Elmegreen}, B.~G., \& {Elmegreen}, D.~M. 2007, \apj, 670, 237

\bibitem[{{Bouwens} {et~al.}(2021){Bouwens}, {Smit}, {Schouws}, {Stefanon},
  {Bowler}, {Endsley}, {Gonzalez}, {Inami}, {Stark}, {Oesch}, {Hodge},
  {Aravena}, {da Cunha}, {Dayal}, {de Looze}, {Ferrara}, {Fudamoto},
  {Graziani}, {Li}, {Nanayakkara}, {Pallotini}, {Schneider}, {Sommovigo},
  {Topping}, {van der Werf}, {Barrufet}, {Hygate}, {Labbe}, {Riechers}, \&
  {Witstok}}]{Bouwens_2021}
{Bouwens}, R.~J., {Smit}, R., {Schouws}, S., {et~al.} 2021, arXiv e-prints,
  arXiv:2106.13719

\bibitem[{{Bovino} {et~al.}(2016){Bovino}, {Grassi}, {Capelo}, {Schleicher}, \&
  {Banerjee}}]{bovino:2016}
{Bovino}, S., {Grassi}, T., {Capelo}, P.~R., {Schleicher}, D. R.~G., \&
  {Banerjee}, R. 2016, \aap, 590, A15

\bibitem[{{Burkert} {et~al.}(2016){Burkert}, {F{\"o}rster Schreiber}, {Genzel},
  {Lang}, {Tacconi}, {Wisnioski}, {Wuyts}, {Bandara}, {Beifiori}, {Bender},
  {Brammer}, {Chan}, {Davies}, {Dekel}, {Fabricius}, {Fossati}, {Kulkarni},
  {Lutz}, {Mendel}, {Momcheva}, {Nelson}, {Naab}, {Renzini}, {Saglia},
  {Sharples}, {Sternberg}, {Wilman}, \& {Wuyts}}]{Burkert_2016}
{Burkert}, A., {F{\"o}rster Schreiber}, N.~M., {Genzel}, R., {et~al.} 2016,
  \apj, 826, 214

\bibitem[{{Ceverino} {et~al.}(2015){Ceverino}, {Dekel}, {Tweed}, \&
  {Primack}}]{Ceverino_2015}
{Ceverino}, D., {Dekel}, A., {Tweed}, D., \& {Primack}, J. 2015, \mnras, 447,
  3291

\bibitem[{{Cimatti} {et~al.}(2019){Cimatti}, {Fraternali}, \&
  {Nipoti}}]{Cimatti_2019}
{Cimatti}, A., {Fraternali}, F., \& {Nipoti}, C. 2019, {Introduction to Galaxy
  Formation and Evolution: From Primordial Gas to Present-Day Galaxies}

\bibitem[{{Cuby} {et~al.}(2019){Cuby}, {Oesch}, {Cooray}, {Rhodes}, {Bremer},
  {Bowler}, {Capak}, {Caputi}, {Castellano}, {Conselice}, {Cuillandre},
  {Dayal}, {Davidzon}, {Dunlop}, {Finkelstein}, {Fontana}, {Kashlinsky},
  {Koopmans}, {Kuijken}, {Le Brun}, {Le F{\`e}vre}, {Mortlock}, {Pello},
  {Pentericci}, {Sahl{\'e}n}, {Schneider}, {Serjeant}, \& {Warren}}]{Cuby_2019}
{Cuby}, J.-G., {Oesch}, P., {Cooray}, A., {et~al.} 2019, \baas, 51, 360

\bibitem[{{Danovich} {et~al.}(2015){Danovich}, {Dekel}, {Hahn}, {Ceverino}, \&
  {Primack}}]{Danovich_2015}
{Danovich}, M., {Dekel}, A., {Hahn}, O., {Ceverino}, D., \& {Primack}, J. 2015,
  \mnras, 449, 2087

\bibitem[{{de Blok} {et~al.}(2014){de Blok}, {Keating}, {Pisano}, {Fraternali},
  {Walter}, {Oosterloo}, {Brinks}, {Bigiel}, \& {Leroy}}]{DeBlok_2014}
{de Blok}, W.~J.~G., {Keating}, K.~M., {Pisano}, D.~J., {et~al.} 2014, \aap,
  569, A68

\bibitem[{{Decataldo} {et~al.}(2019){Decataldo}, {Pallottini}, {Ferrara},
  {Vallini}, \& {Gallerani}}]{Decataldo_2019}
{Decataldo}, D., {Pallottini}, A., {Ferrara}, A., {Vallini}, L., \&
  {Gallerani}, S. 2019, \mnras, 487, 3377

\bibitem[{{Dekel} {et~al.}(2020){Dekel}, {Lapiner}, {Ginzburg}, {Freundlich},
  {Jiang}, {Finish}, {Kretschmer}, {Lin}, {Ceverino}, {Primack}, {Giavalisco},
  \& {Ji}}]{Dekel_2020}
{Dekel}, A., {Lapiner}, S., {Ginzburg}, O., {et~al.} 2020, \mnras, 496, 5372

\bibitem[{{Dekel} {et~al.}(2009){Dekel}, {Sari}, \& {Ceverino}}]{Dekel_2009}
{Dekel}, A., {Sari}, R., \& {Ceverino}, D. 2009, \apj, 703, 785

\bibitem[{{Di Teodoro} \& {Fraternali}(2015)}]{DiTeodoro_2015}
{Di Teodoro}, E.~M. \& {Fraternali}, F. 2015, \mnras, 451, 3021

\bibitem[{{Di Teodoro} \& {Peek}(2021)}]{DiTeodoro_2021}
{Di Teodoro}, E.~M. \& {Peek}, J.~E.~G. 2021, \apj, 923, 220

\bibitem[{{Dubois} {et~al.}(2012){Dubois}, {Pichon}, {Haehnelt}, {Kimm},
  {Slyz}, {Devriendt}, \& {Pogosyan}}]{Dubois_2012}
{Dubois}, Y., {Pichon}, C., {Haehnelt}, M., {et~al.} 2012, \mnras, 423, 3616

\bibitem[{{Duncan} {et~al.}(2019){Duncan}, {Conselice}, {Mundy}, {Bell},
  {Donley}, {Galametz}, {Guo}, {Grogin}, {Hathi}, {Kartaltepe}, {Kocevski},
  {Koekemoer}, {P{\'e}rez-Gonz{\'a}lez}, {Mantha}, {Snyder}, \&
  {Stefanon}}]{Duncan_2019}
{Duncan}, K., {Conselice}, C.~J., {Mundy}, C., {et~al.} 2019, \apj, 876, 110

\bibitem[{{Ejdetj{\"a}rn} {et~al.}(2021){Ejdetj{\"a}rn}, {Agertz},
  {{\"O}stlin}, {Renaud}, \& {Romeo}}]{Ejdetjarn_2021}
{Ejdetj{\"a}rn}, T., {Agertz}, O., {{\"O}stlin}, G., {Renaud}, F., \& {Romeo},
  A.~B. 2021, arXiv e-prints, arXiv:2111.09322

\bibitem[{{Federrath} \& {Klessen}(2013)}]{Federrath_2013}
{Federrath}, C. \& {Klessen}, R.~S. 2013, \apj, 763, 51

\bibitem[{{Federrath} {et~al.}(2017){Federrath}, {Salim}, {Medling}, {Davies},
  {Yuan}, {Bian}, {Groves}, {Ho}, {Sharp}, {Kewley}, {Sweet}, {Richards},
  {Bryant}, {Brough}, {Croom}, {Scott}, {Lawrence}, {Konstantopoulos}, \&
  {Goodwin}}]{Federrath_2017}
{Federrath}, C., {Salim}, D.~M., {Medling}, A.~M., {et~al.} 2017, \mnras, 468,
  3965

\bibitem[{{Ferland} {et~al.}(2017){Ferland}, {Chatzikos}, {Guzm{\'a}n},
  {Lykins}, {van Hoof}, {Williams}, {Abel}, {Badnell}, {Keenan}, {Porter}, \&
  {Stancil}}]{Ferland_2017}
{Ferland}, G.~J., {Chatzikos}, M., {Guzm{\'a}n}, F., {et~al.} 2017, \rmxaa, 53,
  385

\bibitem[{{F{\"o}rster Schreiber} {et~al.}(2009){F{\"o}rster Schreiber},
  {Genzel}, {Bouch{\'e}}, {Cresci}, {Davies}, {Buschkamp}, {Shapiro},
  {Tacconi}, {Hicks}, {Genel}, {Shapley}, {Erb}, {Steidel}, {Lutz},
  {Eisenhauer}, {Gillessen}, {Sternberg}, {Renzini}, {Cimatti}, {Daddi},
  {Kurk}, {Lilly}, {Kong}, {Lehnert}, {Nesvadba}, {Verma}, {McCracken},
  {Arimoto}, {Mignoli}, \& {Onodera}}]{Forster_2009}
{F{\"o}rster Schreiber}, N.~M., {Genzel}, R., {Bouch{\'e}}, N., {et~al.} 2009,
  \apj, 706, 1364

\bibitem[{{Fraternali} {et~al.}(2021){Fraternali}, {Karim}, {Magnelli},
  {G{\'o}mez-Guijarro}, {Jim{\'e}nez-Andrade}, \& {Posses}}]{Fraternali_2021}
{Fraternali}, F., {Karim}, A., {Magnelli}, B., {et~al.} 2021, \aap, 647, A194

\bibitem[{{Fraternali} {et~al.}(2001){Fraternali}, {Oosterloo}, {Sancisi}, \&
  {van Moorsel}}]{Fraternali_2001}
{Fraternali}, F., {Oosterloo}, T., {Sancisi}, R., \& {van Moorsel}, G. 2001,
  \apjl, 562, L47

\bibitem[{{Fraternali} {et~al.}(2002){Fraternali}, {van Moorsel}, {Sancisi}, \&
  {Oosterloo}}]{Fraternali_2002}
{Fraternali}, F., {van Moorsel}, G., {Sancisi}, R., \& {Oosterloo}, T. 2002,
  \aj, 123, 3124

\bibitem[{{Fujimoto} {et~al.}(2021){Fujimoto}, {Oguri}, {Brammer}, {Yoshimura},
  {Laporte}, {Gonz{\'a}lez-L{\'o}pez}, {Caminha}, {Kohno}, {Zitrin}, {Richard},
  {Ouchi}, {Bauer}, {Smail}, {Hatsukade}, {Ono}, {Kokorev}, {Umehata},
  {Schaerer}, {Knudsen}, {Sun}, {Magdis}, {Valentino}, {Ao}, {Toft},
  {Dessauges-Zavadsky}, {Shimasaku}, {Caputi}, {Kusakabe}, {Morokuma-Matsui},
  {Shotaro}, {Egami}, {Lee}, {Rawle}, \& {Espada}}]{Fujimoto_2021}
{Fujimoto}, S., {Oguri}, M., {Brammer}, G., {et~al.} 2021, \apj, 911, 99

\bibitem[{{Fujimoto} {et~al.}(2020){Fujimoto}, {Silverman}, {Bethermin},
  {Ginolfi}, {Jones}, {Le F{\`e}vre}, {Dessauges-Zavadsky}, {Rujopakarn},
  {Faisst}, {Fudamoto}, {Cassata}, {Morselli}, {Maiolino}, {Schaerer}, {Capak},
  {Yan}, {Vallini}, {Toft}, {Loiacono}, {Zamorani}, {Talia}, {Narayanan},
  {Hathi}, {Lemaux}, {Boquien}, {Amorin}, {Ibar}, {Koekemoer},
  {M{\'e}ndez-Hern{\'a}ndez}, {Bardelli}, {Vergani}, {Zucca}, {Romano}, \&
  {Cimatti}}]{Fujimoto_2020}
{Fujimoto}, S., {Silverman}, J.~D., {Bethermin}, M., {et~al.} 2020, \apj, 900,
  1

\bibitem[{{Garc{\'\i}a-Mar{\'\i}n} {et~al.}(2009){Garc{\'\i}a-Mar{\'\i}n},
  {Colina}, \& {Arribas}}]{Garcia_2009}
{Garc{\'\i}a-Mar{\'\i}n}, M., {Colina}, L., \& {Arribas}, S. 2009, \aap, 505,
  1017

\bibitem[{{Glazebrook}(2013)}]{Glazebrook_2013}
{Glazebrook}, K. 2013, \pasa, 30, e056

\bibitem[{{Gon{\c{c}}alves} {et~al.}(2010){Gon{\c{c}}alves}, {Basu-Zych},
  {Overzier}, {Martin}, {Law}, {Schiminovich}, {Wyder}, {Mallery}, {Rich}, \&
  {Heckman}}]{Goncalves_2010}
{Gon{\c{c}}alves}, T.~S., {Basu-Zych}, A., {Overzier}, R., {et~al.} 2010, \apj,
  724, 1373

\bibitem[{{Grassi} {et~al.}(2014){Grassi}, {Bovino}, {Schleicher}, {Prieto},
  {Seifried}, {Simoncini}, \& {Gianturco}}]{Grassi_2014}
{Grassi}, T., {Bovino}, S., {Schleicher}, D.~R.~G., {et~al.} 2014, \mnras, 439,
  2386

\bibitem[{{Hahn} \& {Abel}(2011)}]{Hahn_2011}
{Hahn}, O. \& {Abel}, T. 2011, \mnras, 415, 2101

\bibitem[{{Harikane} {et~al.}(2020){Harikane}, {Ouchi}, {Inoue}, {Matsuoka},
  {Tamura}, {Bakx}, {Fujimoto}, {Moriwaki}, {Ono}, {Nagao}, {Tadaki}, {Kojima},
  {Shibuya}, {Egami}, {Ferrara}, {Gallerani}, {Hashimoto}, {Kohno}, {Matsuda},
  {Matsuo}, {Pallottini}, {Sugahara}, \& {Vallini}}]{Harikane_2020}
{Harikane}, Y., {Ouchi}, M., {Inoue}, A.~K., {et~al.} 2020, \apj, 896, 93

\bibitem[{{Harrison} {et~al.}(2017){Harrison}, {Johnson}, {Swinbank}, {Stott},
  {Bower}, {Smail}, {Tiley}, {Bunker}, {Cirasuolo}, {Sobral}, {Sharples},
  {Best}, {Bureau}, {Jarvis}, \& {Magdis}}]{Harrison_2017}
{Harrison}, C.~M., {Johnson}, H.~L., {Swinbank}, A.~M., {et~al.} 2017, \mnras,
  467, 1965

\bibitem[{{Hashimoto} {et~al.}(2019){Hashimoto}, {Inoue}, {Mawatari}, {Tamura},
  {Matsuo}, {Furusawa}, {Harikane}, {Shibuya}, {Knudsen}, {Kohno}, {Ono},
  {Zackrisson}, {Okamoto}, {Kashikawa}, {Oesch}, {Ouchi}, {Ota}, {Shimizu},
  {Taniguchi}, {Umehata}, \& {Watson}}]{Hashimoto_2019}
{Hashimoto}, T., {Inoue}, A.~K., {Mawatari}, K., {et~al.} 2019, \pasj, 71, 71

\bibitem[{{Herrera} {et~al.}(2012){Herrera}, {Boulanger}, {Nesvadba}, \&
  {Falgarone}}]{Herrera_2012}
{Herrera}, C.~N., {Boulanger}, F., {Nesvadba}, N.~P.~H., \& {Falgarone}, E.
  2012, \aap, 538, L9

\bibitem[{Hunter(2007)}]{matplotlib}
Hunter, J.~D. 2007, Computing in Science Engineering, 9, 90

\bibitem[{{Johnson} {et~al.}(2018){Johnson}, {Harrison}, {Swinbank}, {Tiley},
  {Stott}, {Bower}, {Smail}, {Bunker}, {Sobral}, {Turner}, {Best}, {Bureau},
  {Cirasuolo}, {Jarvis}, {Magdis}, {Sharples}, {Bland-Hawthorn}, {Catinella},
  {Cortese}, {Croom}, {Federrath}, {Glazebrook}, {Sweet}, {Bryant}, {Goodwin},
  {Konstantopoulos}, {Lawrence}, {Medling}, {Owers}, \&
  {Richards}}]{Johnson_2018}
{Johnson}, H.~L., {Harrison}, C.~M., {Swinbank}, A.~M., {et~al.} 2018, \mnras,
  474, 5076

\bibitem[{{Jones} {et~al.}(2021){Jones}, {Vergani}, {Romano}, {Ginolfi},
  {Fudamoto}, {B{\'e}thermin}, {Fujimoto}, {Lemaux}, {Morselli}, {Capak},
  {Cassata}, {Faisst}, {Le F{\`e}vre}, {Schaerer}, {Silverman}, {Yan},
  {Boquien}, {Cimatti}, {Dessauges-Zavadsky}, {Ibar}, {Maiolino}, {Rizzo},
  {Talia}, \& {Zamorani}}]{Jones_2021}
{Jones}, G.~C., {Vergani}, D., {Romano}, M., {et~al.} 2021, \mnras, 507, 3540

\bibitem[{{Jones} {et~al.}(2017){Jones}, {Willott}, {Carilli}, {Ferrara},
  {Wang}, \& {Wagg}}]{Jones_2017}
{Jones}, G.~C., {Willott}, C.~J., {Carilli}, C.~L., {et~al.} 2017, \apj, 845,
  175

\bibitem[{{Kassin} {et~al.}(2012){Kassin}, {Weiner}, {Faber}, {Gardner},
  {Willmer}, {Coil}, {Cooper}, {Devriendt}, {Dutton}, {Guhathakurta}, {Koo},
  {Metevier}, {Noeske}, \& {Primack}}]{Kassin_2012}
{Kassin}, S.~A., {Weiner}, B.~J., {Faber}, S.~M., {et~al.} 2012, \apj, 758, 106

\bibitem[{{Kennicutt}(1998)}]{Kennicutt_1998}
{Kennicutt}, Jr., R.~C. 1998, \apj, 498, 541

\bibitem[{{Kohandel} \& {al., in prep.}(2022)}]{kohandel:2022}
{Kohandel}, M. \& {al., in prep.} 2022, 0, 0

\bibitem[{{Kohandel} {et~al.}(2020){Kohandel}, {Pallottini}, {Ferrara},
  {Carniani}, {Gallerani}, {Vallini}, {Zanella}, \& {Behrens}}]{Kohandel_2020}
{Kohandel}, M., {Pallottini}, A., {Ferrara}, A., {et~al.} 2020, \mnras, 499,
  1250

\bibitem[{{Kohandel} {et~al.}(2019){Kohandel}, {Pallottini}, {Ferrara},
  {Zanella}, {Behrens}, {Carniani}, {Gallerani}, \& {Vallini}}]{Kohandel_2019}
{Kohandel}, M., {Pallottini}, A., {Ferrara}, A., {et~al.} 2019, \mnras, 487,
  3007

\bibitem[{{Krajnovi{\'c}} {et~al.}(2006){Krajnovi{\'c}}, {Cappellari}, {de
  Zeeuw}, \& {Copin}}]{Krajnovic_2006}
{Krajnovi{\'c}}, D., {Cappellari}, M., {de Zeeuw}, P.~T., \& {Copin}, Y. 2006,
  \mnras, 366, 787

\bibitem[{{Kretschmer} {et~al.}(2021){Kretschmer}, {Dekel}, \&
  {Teyssier}}]{Kretschmer_2021}
{Kretschmer}, M., {Dekel}, A., \& {Teyssier}, R. 2021, \mnras
  [\eprint[arXiv]{2103.06882}]

\bibitem[{{Krumholz} {et~al.}(2018){Krumholz}, {Burkhart}, {Forbes}, \&
  {Crocker}}]{Krumholz_2018}
{Krumholz}, M.~R., {Burkhart}, B., {Forbes}, J.~C., \& {Crocker}, R.~M. 2018,
  \mnras, 477, 2716

\bibitem[{{Law} {et~al.}(2009){Law}, {Steidel}, {Erb}, {Larkin}, {Pettini},
  {Shapley}, \& {Wright}}]{Law_2009}
{Law}, D.~R., {Steidel}, C.~C., {Erb}, D.~K., {et~al.} 2009, \apj, 697, 2057

\bibitem[{{Le F{\`e}vre} {et~al.}(2020){Le F{\`e}vre}, {B{\'e}thermin},
  {Faisst}, {Jones}, {Capak}, {Cassata}, {Silverman}, {Schaerer}, {Yan},
  {Amorin}, {Bardelli}, {Boquien}, {Cimatti}, {Dessauges-Zavadsky},
  {Giavalisco}, {Hathi}, {Fudamoto}, {Fujimoto}, {Ginolfi}, {Gruppioni},
  {Hemmati}, {Ibar}, {Koekemoer}, {Khusanova}, {Lagache}, {Lemaux}, {Loiacono},
  {Maiolino}, {Mancini}, {Narayanan}, {Morselli}, {M{\'e}ndez-Hern{\`a}ndez},
  {Oesch}, {Pozzi}, {Romano}, {Riechers}, {Scoville}, {Talia}, {Tasca},
  {Thomas}, {Toft}, {Vallini}, {Vergani}, {Walter}, {Zamorani}, \&
  {Zucca}}]{LeFevre_2020}
{Le F{\`e}vre}, O., {B{\'e}thermin}, M., {Faisst}, A., {et~al.} 2020, \aap,
  643, A1

\bibitem[{{Lelli} {et~al.}(2021){Lelli}, {Di Teodoro}, {Fraternali}, {Man},
  {Zhang}, {De Breuck}, {Davis}, \& {Maiolino}}]{Lelli_2021}
{Lelli}, F., {Di Teodoro}, E.~M., {Fraternali}, F., {et~al.} 2021, Science,
  371, 713

\bibitem[{{Mantha} {et~al.}(2018){Mantha}, {McIntosh}, {Brennan}, {Ferguson},
  {Kodra}, {Newman}, {Rafelski}, {Somerville}, {Conselice}, {Cook}, {Hathi},
  {Koo}, {Lotz}, {Simmons}, {Straughn}, {Snyder}, {Wuyts}, {Bell}, {Dekel},
  {Kartaltepe}, {Kocevski}, {Koekemoer}, {Lee}, {Lucas}, {Pacifici}, {Peth},
  {Barro}, {Dahlen}, {Finkelstein}, {Fontana}, {Galametz}, {Grogin}, {Guo},
  {Mobasher}, {Nayyeri}, {P{\'e}rez-Gonz{\'a}lez}, {Pforr}, {Santini},
  {Stefanon}, \& {Wiklind}}]{Mantha_2018}
{Mantha}, K.~B., {McIntosh}, D.~H., {Brennan}, R., {et~al.} 2018, \mnras, 475,
  1549

\bibitem[{{Mason} {et~al.}(2015){Mason}, {Trenti}, \& {Treu}}]{Mason_2015}
{Mason}, C.~A., {Trenti}, M., \& {Treu}, T. 2015, \apj, 813, 21

\bibitem[{{McMullin} {et~al.}(2007){McMullin}, {Waters}, {Schiebel}, {Young},
  \& {Golap}}]{McMullin_2007}
{McMullin}, J.~P., {Waters}, B., {Schiebel}, D., {Young}, W., \& {Golap}, K.
  2007, in Astronomical Society of the Pacific Conference Series, Vol. 376,
  Astronomical Data Analysis Software and Systems XVI, ed. R.~A. {Shaw},
  F.~{Hill}, \& D.~J. {Bell}, 127

\bibitem[{{Neeleman} {et~al.}(2020){Neeleman}, {Prochaska}, {Kanekar}, \&
  {Rafelski}}]{Neeleman_2020}
{Neeleman}, M., {Prochaska}, J.~X., {Kanekar}, N., \& {Rafelski}, M. 2020,
  \nat, 581, 269

\bibitem[{{Pallottini} {et~al.}(2019){Pallottini}, {Ferrara}, {Decataldo},
  {Gallerani}, {Vallini}, {Carniani}, {Behrens}, {Kohandel}, \&
  {Salvadori}}]{Pallottini_2019}
{Pallottini}, A., {Ferrara}, A., {Decataldo}, D., {et~al.} 2019, \mnras, 487,
  1689

\bibitem[{{Pallottini} {et~al.}(2022){Pallottini}, {Ferrara}, {Gallerani},
  {Behrens}, {Kohandel}, {Carniani}, {Vallini}, {Salvadori}, {Gelli},
  {Sommovigo}, {D'Odorico}, {Di Mascia}, \& {Pizzati}}]{Pallottini_2022}
{Pallottini}, A., {Ferrara}, A., {Gallerani}, S., {et~al.} 2022, arXiv
  e-prints, arXiv:2201.02636

\bibitem[{{Pallottini} {et~al.}(2017{\natexlab{a}}){Pallottini}, {Ferrara},
  {Gallerani}, {Vallini}, {Maiolino}, \& {Salvadori}}]{Pallottini_2017dahlia}
{Pallottini}, A., {Ferrara}, A., {Gallerani}, S., {et~al.} 2017{\natexlab{a}},
  \mnras, 465, 2540

\bibitem[{{Pallottini} {et~al.}(2017{\natexlab{b}}){Pallottini}, {Ferrara},
  {Gallerani}, {Vallini}, {Maiolino}, \& {Salvadori}}]{Pallottini_2017}
{Pallottini}, A., {Ferrara}, A., {Gallerani}, S., {et~al.} 2017{\natexlab{b}},
  \mnras, 465, 2540

\bibitem[{{Pedregosa} {et~al.}(2012){Pedregosa}, {Varoquaux}, {Gramfort},
  {Michel}, {Thirion}, {Grisel}, {Blondel}, {M{\"u}ller}, {Nothman}, {Louppe},
  {Prettenhofer}, {Weiss}, {Dubourg}, {Vanderplas}, {Passos}, {Cournapeau},
  {Brucher}, {Perrot}, \& {Duchesnay}}]{pedregosa:2012}
{Pedregosa}, F., {Varoquaux}, G., {Gramfort}, A., {et~al.} 2012, arXiv
  e-prints, arXiv:1201.0490

\bibitem[{{Pillepich} {et~al.}(2019){Pillepich}, {Nelson}, {Springel},
  {Pakmor}, {Torrey}, {Weinberger}, {Vogelsberger}, {Marinacci}, {Genel}, {van
  der Wel}, \& {Hernquist}}]{Pillepich_2019}
{Pillepich}, A., {Nelson}, D., {Springel}, V., {et~al.} 2019, \mnras, 490, 3196

\bibitem[{{Pontzen} {et~al.}(2013){Pontzen}, {Rovskar}, {Stinson}, {Woods},
  {Reed}, {Coles}, \& {Quinn}}]{pynbody}
{Pontzen}, A., {Rovskar}, R., {Stinson}, G.~S., {et~al.} 2013, {pynbody:
  Astrophysics Simulation Analysis for Python}, astrophysics Source Code
  Library, ascl:1305.002

\bibitem[{{Ramos Almeida} {et~al.}(2021){Ramos Almeida}, {Bischetti},
  {Garcia-Burillo}, {Alonso-Herrero}, {Audibert}, {Cicone}, {Feruglio},
  {Tadhunter}, {Pierce}, {Pereira-Santaella}, \& {Bessiere}}]{Almeida_2021}
{Ramos Almeida}, C., {Bischetti}, M., {Garcia-Burillo}, S., {et~al.} 2021,
  arXiv e-prints, arXiv:2111.13578

\bibitem[{{Rizzo} {et~al.}(2021){Rizzo}, {Vegetti}, {Fraternali}, {Stacey}, \&
  {Powell}}]{Rizzo_2021}
{Rizzo}, F., {Vegetti}, S., {Fraternali}, F., {Stacey}, H.~R., \& {Powell}, D.
  2021, \mnras, 507, 3952

\bibitem[{{Rizzo} {et~al.}(2020){Rizzo}, {Vegetti}, {Powell}, {Fraternali},
  {McKean}, {Stacey}, \& {White}}]{Rizzo_2020}
{Rizzo}, F., {Vegetti}, S., {Powell}, D., {et~al.} 2020, \nat, 584, 201

\bibitem[{{Rodrigues} {et~al.}(2017){Rodrigues}, {Hammer}, {Flores}, {Puech},
  \& {Athanassoula}}]{Rodrigues_2017}
{Rodrigues}, M., {Hammer}, F., {Flores}, H., {Puech}, M., \& {Athanassoula}, E.
  2017, \mnras, 465, 1157

\bibitem[{{Rogstad} {et~al.}(1974){Rogstad}, {Lockhart}, \&
  {Wright}}]{Rogstad_1974}
{Rogstad}, D.~H., {Lockhart}, I.~A., \& {Wright}, M.~C.~H. 1974, \apj, 193, 309

\bibitem[{{Romano} {et~al.}(2021){Romano}, {Cassata}, {Morselli}, {Jones},
  {Ginolfi}, {Zanella}, {B{\'e}thermin}, {Capak}, {Faisst}, {Le F{\`e}vre},
  {Schaerer}, {Silverman}, {Yan}, {Bardelli}, {Boquien}, {Cimatti},
  {Dessauges-Zavadsky}, {Enia}, {Fujimoto}, {Gruppioni}, {Hathi}, {Ibar},
  {Koekemoer}, {Lemaux}, {Rodighiero}, {Vergani}, {Zamorani}, \&
  {Zucca}}]{Romano_2021}
{Romano}, M., {Cassata}, P., {Morselli}, L., {et~al.} 2021, \aap, 653, A111

\bibitem[{{Rosdahl} {et~al.}(2013){Rosdahl}, {Blaizot}, {Aubert}, {Stranex}, \&
  {Teyssier}}]{Rosdahl_2013}
{Rosdahl}, J., {Blaizot}, J., {Aubert}, D., {Stranex}, T., \& {Teyssier}, R.
  2013, \mnras, 436, 2188

\bibitem[{{Schmidt}(1959)}]{Schmidt_1959}
{Schmidt}, M. 1959, \apj, 129, 243

\bibitem[{{Shapiro} {et~al.}(2008){Shapiro}, {Genzel}, {F{\"o}rster Schreiber},
  {Tacconi}, {Bouch{\'e}}, {Cresci}, {Davies}, {Eisenhauer}, {Johansson},
  {Krajnovi{\'c}}, {Lutz}, {Naab}, {Arimoto}, {Arribas}, {Cimatti}, {Colina},
  {Daddi}, {Daigle}, {Erb}, {Hernandez}, {Kong}, {Mignoli}, {Onodera},
  {Renzini}, {Shapley}, \& {Steidel}}]{Shapiro_2008}
{Shapiro}, K.~L., {Genzel}, R., {F{\"o}rster Schreiber}, N.~M., {et~al.} 2008,
  \apj, 682, 231

\bibitem[{{Simons} {et~al.}(2019){Simons}, {Kassin}, {Snyder}, {Primack},
  {Ceverino}, {Dekel}, {Hayward}, {Mandelker}, {Mantha}, {Pacifici}, {de la
  Vega}, \& {Wang}}]{Simons_2019}
{Simons}, R.~C., {Kassin}, S.~A., {Snyder}, G.~F., {et~al.} 2019, \apj, 874, 59

\bibitem[{{Smit} {et~al.}(2018){Smit}, {Bouwens}, {Carniani}, {Oesch},
  {Labb{\'e}}, {Illingworth}, {van der Werf}, {Bradley}, {Gonzalez}, {Hodge},
  {Holwerda}, {Maiolino}, \& {Zheng}}]{Smit_2018}
{Smit}, R., {Bouwens}, R.~J., {Carniani}, S., {et~al.} 2018, \nat, 553, 178

\bibitem[{{Somerville} \& {Dav{\'e}}(2015)}]{Somerville_2015}
{Somerville}, R.~S. \& {Dav{\'e}}, R. 2015, \araa, 53, 51

\bibitem[{{Stott} {et~al.}(2016){Stott}, {Swinbank}, {Johnson}, {Tiley},
  {Magdis}, {Bower}, {Bunker}, {Bureau}, {Harrison}, {Jarvis}, {Sharples},
  {Smail}, {Sobral}, {Best}, \& {Cirasuolo}}]{Stott_2016}
{Stott}, J.~P., {Swinbank}, A.~M., {Johnson}, H.~L., {et~al.} 2016, \mnras,
  457, 1888

\bibitem[{{Swaters}(1999)}]{Swaters_1999}
{Swaters}, R.~A. 1999, PhD thesis, -

\bibitem[{{Swinbank} {et~al.}(2012){Swinbank}, {Smail}, {Sobral}, {Theuns},
  {Best}, \& {Geach}}]{Swinbank_2012}
{Swinbank}, A.~M., {Smail}, I., {Sobral}, D., {et~al.} 2012, \apj, 760, 130

\bibitem[{{Tamfal} {et~al.}(2021){Tamfal}, {Mayer}, {Quinn}, {Babul}, {Madau},
  {Capelo}, {Shen}, \& {Staub}}]{Tamfal_2021}
{Tamfal}, T., {Mayer}, L., {Quinn}, T.~R., {et~al.} 2021, arXiv e-prints,
  arXiv:2106.11981

\bibitem[{{Teyssier}(2002)}]{Teyssier_2002}
{Teyssier}, R. 2002, \aap, 385, 337

\bibitem[{{Turner} {et~al.}(2017){Turner}, {Cirasuolo}, {Harrison}, {McLure},
  {Dunlop}, {Swinbank}, {Johnson}, {Sobral}, {Matthee}, \&
  {Sharples}}]{Turner_2017}
{Turner}, O.~J., {Cirasuolo}, M., {Harrison}, C.~M., {et~al.} 2017, \mnras,
  471, 1280

\bibitem[{{Vallini} {et~al.}(2018){Vallini}, {Pallottini}, {Ferrara},
  {Gallerani}, {Sobacchi}, \& {Behrens}}]{Vallini_2018}
{Vallini}, L., {Pallottini}, A., {Ferrara}, A., {et~al.} 2018, \mnras, 473, 271

\bibitem[{van~der Walt {et~al.}(2011)van~der Walt, Colbert, \&
  Varoquaux}]{numpy}
van~der Walt, S., Colbert, S.~C., \& Varoquaux, G. 2011, Computing in Science
  Engineering, 13, 22

\bibitem[{Van~Rossum \& de~Boer(1991)}]{python2}
Van~Rossum, G. \& de~Boer, J. 1991, CWI Quarterly, 4, 283

\bibitem[{Van~Rossum \& Drake(2009)}]{python3}
Van~Rossum, G. \& Drake, F.~L. 2009, Python 3 Reference Manual (Scotts Valley,
  CA: CreateSpace)

\bibitem[{{Varidel} {et~al.}(2016){Varidel}, {Pracy}, {Croom}, {Owers}, \&
  {Sadler}}]{Varidel_2016}
{Varidel}, M., {Pracy}, M., {Croom}, S., {Owers}, M.~S., \& {Sadler}, E. 2016,
  \pasa, 33, e006

\bibitem[{{Virtanen} {et~al.}(2020){Virtanen}, {Gommers}, {Oliphant},
  {Haberland}, {Reddy}, {Cournapeau}, {Burovski}, {Peterson}, {Weckesser},
  {Bright}, {van der Walt}, {Brett}, {Wilson}, {Millman}, {Mayorov}, {Nelson},
  {Jones}, {Kern}, {Larson}, {Carey}, {Polat}, {Feng}, {Moore}, {VanderPlas},
  {Laxalde}, {Perktold}, {Cimrman}, {Henriksen}, {Quintero}, {Harris},
  {Archibald}, {Ribeiro}, {Pedregosa}, {van Mulbregt}, \& {SciPy 1. 0
  Contributors}}]{scipy}
{Virtanen}, P., {Gommers}, R., {Oliphant}, T.~E., {et~al.} 2020, Nature
  Methods, 17, 261

\bibitem[{{Vogelsberger} {et~al.}(2020){Vogelsberger}, {Nelson}, {Pillepich},
  {Shen}, {Marinacci}, {Springel}, {Pakmor}, {Tacchella}, {Weinberger},
  {Torrey}, \& {Hernquist}}]{Vogelsberger_2020}
{Vogelsberger}, M., {Nelson}, D., {Pillepich}, A., {et~al.} 2020, \mnras, 492,
  5167

\bibitem[{{Wisnioski} {et~al.}(2019){Wisnioski}, {F{\"o}rster Schreiber},
  {Fossati}, {Mendel}, {Wilman}, {Genzel}, {Bender}, {Wuyts}, {Davies},
  {{\"U}bler}, {Bandara}, {Beifiori}, {Belli}, {Brammer}, {Chan}, {Davies},
  {Fabricius}, {Galametz}, {Lang}, {Lutz}, {Nelson}, {Momcheva}, {Price},
  {Rosario}, {Saglia}, {Seitz}, {Shimizu}, {Tacconi}, {Tadaki}, {van Dokkum},
  \& {Wuyts}}]{Wisnioski_2018}
{Wisnioski}, E., {F{\"o}rster Schreiber}, N.~M., {Fossati}, M., {et~al.} 2019,
  \apj, 886, 124

\bibitem[{{Wisnioski} {et~al.}(2015){Wisnioski}, {F{\"o}rster Schreiber},
  {Wuyts}, {Wuyts}, {Bandara}, {Wilman}, {Genzel}, {Bender}, {Davies},
  {Fossati}, {Lang}, {Mendel}, {Beifiori}, {Brammer}, {Chan}, {Fabricius},
  {Fudamoto}, {Kulkarni}, {Kurk}, {Lutz}, {Nelson}, {Momcheva}, {Rosario},
  {Saglia}, {Seitz}, {Tacconi}, \& {van Dokkum}}]{Wisnioski_2015}
{Wisnioski}, E., {F{\"o}rster Schreiber}, N.~M., {Wuyts}, S., {et~al.} 2015,
  \apj, 799, 209

\bibitem[{{Wolfire} {et~al.}(2022){Wolfire}, {Vallini}, \&
  {Chevance}}]{Wolfire_2022}
{Wolfire}, M.~G., {Vallini}, L., \& {Chevance}, M. 2022, arXiv e-prints,
  arXiv:2202.05867

\bibitem[{{Wootten} \& {Thompson}(2009)}]{Wootten_2009}
{Wootten}, A. \& {Thompson}, A.~R. 2009, IEEE Proceedings, 97, 1463

\bibitem[{{Yttergren} {et~al.}(2021){Yttergren}, {Misquitta},
  {S{\'a}nchez-Monge}, {Valencia-S}, {Eckart}, {Zensus}, \&
  {Peitl-Thiesen}}]{Yttergren_2021}
{Yttergren}, M., {Misquitta}, P., {S{\'a}nchez-Monge}, {\'A}., {et~al.} 2021,
  \aap, 656, A83

\bibitem[{{Zolotov} {et~al.}(2015){Zolotov}, {Dekel}, {Mandelker}, {Tweed},
  {Inoue}, {DeGraf}, {Ceverino}, {Primack}, {Barro}, \& {Faber}}]{Zolotov_2015}
{Zolotov}, A., {Dekel}, A., {Mandelker}, N., {et~al.} 2015, \mnras, 450, 2327

\end{thebibliography}


\begin{appendix}
\section{Kinematic fitting and assumptions}\label{appendix:kinematic_assumptions}

Before proceeding with the \bba\ fitting of the mock cubes (see Sect. \ref{sec:barolo}), we assumed: 
\begin{itemize}
\item For the high- and medium-resolution observations, the separation between rings is set equal to the nominal resolution of the observations (see Sect. \ref{sec:mock_data}). This represents a small oversampling, on average less than 10$\%$ of the size of the beam along the major axis $B_{\mathrm{maj}}$, and it allows to recover of almost independent values of the rotation velocities and velocity dispersions.\\
For most of the low-resolution observations, this assumption on ring separation can not be made because few resolution elements cover the [CII] emission. Since a minimum of two rings is needed for the kinematic fitting, we set this separation equal to $0.08"$ in these cases. Given the sizes of the beams, this choice makes the kinematic parameters extracted from the two rings not fully independent.\\
In Table \ref{table:nire}, we list the number of independent resolution elements $N_{\mathrm{IRE}}$ for the mock data fitted with { \bba\ } (see Sect. \ref{sec:barolo}).
\item We fix the center of the kinematic models to the flux-weighted average position of the moment-0 map when only one single galaxy is visible, that is, for the disk galaxies Opuntia, Lantana, Petunia, Freesia and for the low-resolution merging systems at some inclinations (e.g., Fuchsia). When the two interacting galaxies are visible, such as for the high-resolution observations, we fix the center to the flux-weighted average positions of the two nuclei.
\item When both the geometrical (e.g., $i$, $PA$) and kinematic parameters are free parameters, \bba\ applies a 2-step fitting process. First, it fits the kinematic and geometric parameters together. Then, after interpolating the geometric parameters and fixing them using a regularizing functional form, it fits the kinematic parameters. We use the default regularizing functional form for the high-resolution observations and a constant value for the medium and low-resolution observations.
\item We remove the surface brightness from the list of the free parameters. We use the parameter \texttt{NORM = AZIM} so that the model is normalized to the azimuthally-averaged flux of the data in each ring.
\end{itemize} 

     \begin{table*}\centering
    \caption{Number of independent resolution elements, $N_{\mathrm{IRE}}$, per galaxy side along the semi-major axis computed as described in Appendix \ref{appendix:kinematic_assumptions}.}
    \ra{1.3}
    \begin{tabular}{@{}rrcccccccc@{}}\toprule
    \multicolumn{1}{c}{Galaxy} & \multicolumn{1}{c}{$i$} & \phantom{abc} & \multicolumn{1}{c}{Ideal Mock} &
    \phantom{abc} & \multicolumn{5}{c}{ALMA Mocks} \\ \cmidrule{4-4} \cmidrule{6-10}
        &&& High SN  &&  High SN &  \multicolumn{2}{c}{High SN}  & \multicolumn{2}{c}{Low SN }\\ 
        &&& High res &&  High res &  \multicolumn{2}{c}{Medium \& low res}  & \multicolumn{2}{c}{High \& medium res}\\ \midrule\midrule
            
        Adenia    & 30    &&  8 && 5.2 & 2 & 1.5  \\
                  & 60    &&  7.2  && 7 & 3 & 1.2 & 7 & 3.2 \\
                  & 80    &&  5.3 && 4.4 & 2 & 1.5\\
            \noalign{\smallskip}
        Fuchsia   & 30    && 7.2 && 6.4 & 3.3 & 1.3\\
                  & 60    && 7.2 && 4.8 & 2.6 & 1.6 &  6.7 & 3\\
                  & 80    && 5.4 && 4.8 & 2.4 & 1.3\\
        \noalign{\smallskip}
        Freesia   & 30    && 10.7 && 9 &  4 & 2 \\
                  & 60    && 10.7 && 8 & 4 & 2.2 & 6.4 & 4\\
                  & 80    && 10.7 && 9 &  4 & 2  \\
        \noalign{\smallskip}
        Opuntia   & 30    && 7 && 7 & 3 & 1.2\\
                  & 60    && 8 && 7 & 3 & 1.6 & 5 & 3\\
                  & 80    && 7 && 7 & 3 & 1.6\\
        \noalign{\smallskip}
        Lantana   & 30    && 7 && 6.8 & 2.8 & 1.5 \\
                  & 60    && 8 && 7.7 & 2.8 & 1.5 & 5.5 & 2.4\\
                  & 80    && 8 && 7.6 & 2.8 & 1.8\\
        \noalign{\smallskip}
        Petunia   & 30       && 6.2 && 6 & 2 & 1.1\\
                  & 60       && 6.2 && 5 & 2.3 & 1.1 & 4.3 & 2.3 \\
                  & 80       && 6.2 && 6 & 2 & 1.1\\
        \noalign{\smallskip}
    \bottomrule
    \end{tabular}
    \end{table*} \label{table:nire}
   
\section{A-posteriori beam-smearing correction}\label{appendix:gradient}

In Sect. \ref{sec:kinematic_outputs}, we compared the velocity dispersions obtained with a 3D forward-modeling technique with the ones inferred with a 2D a-posteriori beam-smearing correction. For the latter, we followed the methods proposed by \citet{Swinbank_2012, Varidel_2016, Stott_2016, Federrath_2017}, namely the beam-smearing corrected velocity dispersion is obtained after linearly subtracting the velocity gradient from the observed velocity dispersions. The velocity gradient across the synthesized beam, that is, on average, a factor of 5 larger than the pixel size, is computed for a given pixel with coordinate indices ($i, j$) as
\begin{equation}
   V_{\mathrm{grad}} = \frac{\sqrt{|V(i+2, j)-V(i-2, j)|^2 + |V(i, j+2)-V(i, j-2)|^2}}{5}.
\end{equation}

\section{Low-z galaxies at low angular resolution}\label{appendix:lowz}

NGC2403 is a nearby spiral galaxy that is part of the M81 group. The computation of the PV parameters described in Sect. \ref{sec:new_par} is obtained after modeling its kinematics with \bba. We used single dish HI observations from the Green Bank Telescope \citep{DeBlok_2014, DiTeodoro_2015} with a synthesized beam size of 29.7" $\times$ 29.3", allowing a sampling of the semi-major axis with 3 independent resolution elements.

The Antennae or NGC4038/9 are two spiral galaxies and the nearest example of a major galaxy merger. In Sect. \ref{sec:new_par}, we used the ALMA Science Verification data targeting their CO(3-2) emission line \citep[e.g.,][]{Herrera_2012}. Since we wanted to apply {\bba} and PVsplit on medium/low-angular resolution data, we convolved the ALMA high-angular resolution datacube (beam size of 1.2" $\times$ 0.6") with a Gaussian beam of 122". As a result, similarly to the mock data of the simulated merging galaxies discussed in this work, the two interacting galaxies are indistinguishable and we fit the kinematics of the system with 2 independent resolution elements along the semi-major axis.

\section{PV parameters and angular resolution}

\begin{figure*}
   \centering
        \includegraphics[scale=0.9]{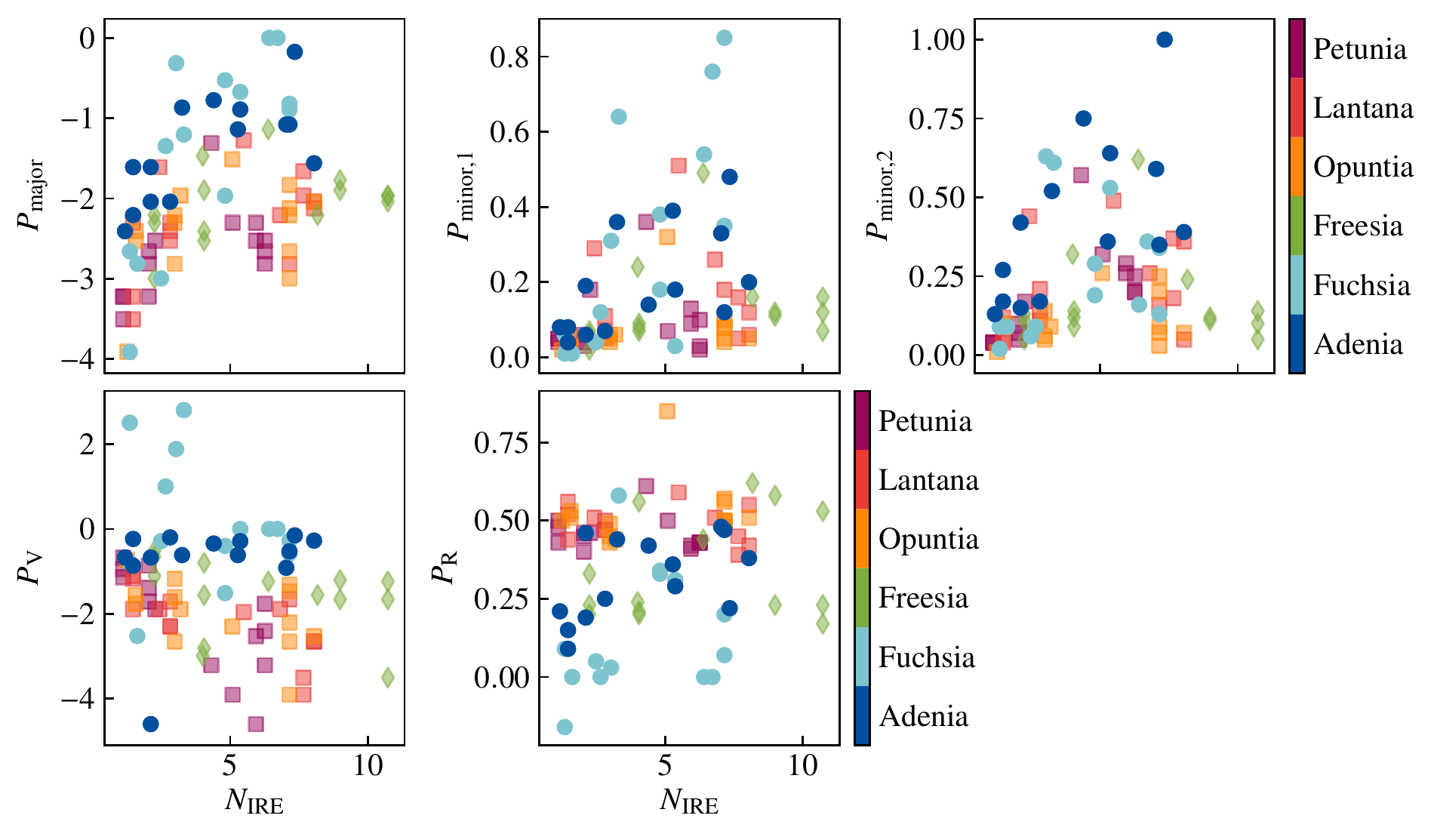}
        \caption{PV parameters described in Sect. \ref{sec:new_par} as a function of the independent resolution elements along the semi-major axis (Appendix \ref{appendix:kinematic_assumptions} for details).
        \label{fig:pvN}%
        }
    \end{figure*}

In Fig. \ref{fig:pvN}, we show the PV parameters as a function of $N_{\mathrm{IRE}}$ for the 84 mock data a) - d) described in Sect. \ref{sec:mock_data}.

\end{appendix}

\end{document}